\documentclass[singlecolumn,epjc3]{svjour3}
\smartqed  
\RequirePackage{graphicx}
\journalname{Eur. Phys. J. C}

\begin{document}

\title{A family of charged compact objects with anisotropic pressure}

\author{S.K. Maurya\thanksref{e1,addr1}\and M. Govender \thanksref{e2,addr2}}

\thankstext{e1}{e-mail: sunil@unizwa.edu.om}
\thankstext{e2}{e-mail: megandhreng@dut.ac.za}

\institute{Department of Mathematical and Physical Sciences,
College of Arts and Science, University of Nizwa, Nizwa, Sultanate
of Oman\label{addr1}
\and Department of Mathematics, Durban University of Technology, Durban, 4000, South Africa. \label{addr2}}

\date{Received: date / Accepted: date}

\maketitle

\begin{abstract}
Utilizing an ansatz developed by Maurya and co-workers we present a class of exact solutions of the Einstein-Maxwell field equations describing a spherically symmetric compact object. A detailed physical analysis of these solutions in terms of stability, compactness and regularity indicate that these solutions may be used to model strange star candidates. In particular, we model the strange star candidate Her X-1 and show that our solution conforms to observational data to an excellent degree of accuracy. An interesting and novel phenomenon which arises in this model is the fact that the relative difference between the electromagnetic force and the force due to pressure anisotropy changes sign within the stellar interior. This may be a an additional mechanism required for stability against cracking of the stellar object.
\end{abstract}


\maketitle

\section{Introduction}

General relativity (GR) has proved to be extremely fruitful in describing the physical universe and the structures contained therein. Over and above in contributing to our understanding of astrophysical bodies and the evolution of the Universe, GR is quintessential in understanding the nature of gravity and it's behaviour in the presence of extremely dense sources as well as in higher dimensions\cite{mtw,hawkis}. With the discovery and observations of ultra-compact objects such as pulsars, neutron stars and black holes, the search for exact solutions of the Einstein field equations has moved away from mere mathematical excursions into the realm of modeling physical objects based on observational data\cite{bhs}. It's just over a century since the first solution of the Einstein field equations describing a self-gravitating, bounded object was first obtained by Karl Schwarzschild\cite{ks}. The Schwarzschild interior solution describes a uniform density sphere and is a first approximation in describing the gravitational field of a static, spherically symmetric object. This solution is highly idealised and propagation speeds within the object exceed the speed of light thus rendering the model noncausal. Since the pioneering effort of Schwarzschild in obtaining exact solutions of the Einstein field equations describing the gravitational field of a bounded mass, there has been a massive drive to obtain realistic solutions describing self-gravitating objects. An indepth study of the available exact solutions in the literature indicate that many of them fall short of describing physically realizable stellar structures\cite{lake1}. Many of these solutions are only valid in some region of the object, other solutions display unphysical behaviour in the density and pressure profiles while many stellar models are unstable against radial perturbations. The wide body of exact solutions which currently exist were obtained through various assumptions on the spacetime geometry, matter content or both\cite{kramer}. Spherical symmetry is the natural assumption to make when modeling static stars. There is however, more freedom in choosing the matter content of the stellar fluid. In the past researchers have worked with perfect fluids, charged interiors, pressure anisotropy, bulk viscosity and scalar fields. More recently, motivated by developments in cosmology, modelling of stellar structures have included dark energy, dark matter and phantom energy\cite{rah1,lobo1}.

Departure from spherical symmetry has been utilised in modeling stars. The Vaidya-Tikekar (VT) superdense stellar model incorporates a spheroidal geometry for the interior of the star\cite{Tikekar90}. The VT model has been shown to approximate the behaviour of neutron stars to a very good approximation\cite{meganvt}. Tikekar and co-workers have successfully modeled stars with paraboidal symmetry. The spheroidal parameter which appears in the gravitational potential measures the deviation from spherical symmetry. Work on rotating stars utilise an axis-symmetric metric to describe the stellar interior\cite{herrera1,herrera2}. Recently there has been a surge in obtaining exact solutions of the Einstein field equations via embedding\cite{sgov,npg,k1,k2,k3,k4,k5,k6,k7}. In 1947 Karmarkar obtained a restriction which is a necessary condition for embedding a spherically symmetric spacetime in four dimensions into a flat five dimensional spacetime. In general, an $n$-dimensional Riemannian spacetime is said to be of class $p$ if it can be embedded into a flat space of dimension $n + p$ \cite{kar4}. The Karmarkar condition relates to class 1 spacetimes. Pandey and Sharma later showed that the Karmarakar condition is only a necessary condition for a spacetime to be of class 1\cite{pandey1}. A further requirement has to be imposed for sufficiency of the Karmarkar condition. The derivation of the Karmarkar condition is purely geometric in nature which gives a relationship between the two gravitational potentials. This is useful because in order to obtain a complete description of the gravitational behaviour of the model one needs to specify one of the metric functions and the other is obtained via the Karmarakar condition. It is also interesting to note that the Karmarkar condition together with the assumption of pressure isotropy picks out the interior Schwarzschild solution as the {\em only} bounded matter configuration with vanishing pressure anisotropy. It follows that if the interior metric of a bounded sphere is of  class 1, then the matter content is necessarily anisotropic or charged, with the Schwarzschild interior solution being the only exception.

In this paper we present a model of a spherically symmetric, charged object obtained by embedding a spherically symmetric static metric in Schwarzschild coordinates into a five dimensional flat space. The pressure within the fluid distribution is anisotropic. The resulting condition arising from the embedding reduces the problem of finding an exact solution to the Einstein-Maxwell equations to specifying one of the gravitational potentials and the behaviour of the electric field intensity.

\section{The Einstein-Maxwell Field Equations for Charged Anisotropic Matter Distribution}

The line element describing the interior of a static, spherically symmetric matter distribution is given in
Schwarzschild coordinates \cite{Oppenheimer1939,Tolman1939}
$x^i=(r, \theta, \phi, t)$ as follows:
\begin{equation}
ds^2 = - e^{\lambda(r)} dr^2 - r^2(d\theta^2 + \sin^2\theta d\phi^2) + e^{\nu(r)} dt^2.\label{metric1}
\end{equation}
where we seek the radial dependence of the potentials $\lambda$ and $\nu$.

In this work we study charged compact objects within the framework of classical general relativity. The Einstein-Maxwell field equations relating the spacetime geometry to the matter content are
\begin{equation}
 {R^i}_j - \frac{1}{2} R\, {g^i}_j = \kappa ({T^i}_j + {E^i}_j).\label{field}
\end{equation}
where $\kappa = 8\pi$ is the Einstein coupling constant. We use geometrized units in which $G=1=c$
with $G$ and $c$ being the
Newtonian gravitational constant and speed of photons in vacuum, respectively.

We assume that the radial and tangential stresses within the interior matter distribution are unequal thus implying that the matter within the star is locally anisotropic. The energy-momentum tensor of fluid the distribution and electromagnetic
field are defined respectively as \cite{Dionysiou}
\begin{equation}
{T^i}_j = [(\rho + p_t)v^iv_j - p_t{\delta^i}_j + (p_r - p_t) \theta^i \theta_j],\label{matter}
\end{equation}

\begin{equation}
{E^i}_j = \frac{1}{4}(-F^{im}F_{jm} + \frac{1}{4}{\delta^i}_jF^{mn}F_{mn}).\label{electric}
\end{equation}
where $v^i$ is the four-velocity, $v^i=e^{\nu(r)/2}{\delta^i}_4$, $\theta^i$ is a unit space-like
vector in the radial direction, $\theta^i =
e^{\lambda(r)/2}{\delta^i}_1$, $\rho $ is the energy density,
$p_r$ is the radial pressure and and $p_t$ is the tangential pressure. The components for ${T^i}_j$ and ${E^i}_j$
are defined respectively as:
\begin{equation}
 {T^1}_1=-p_r,\, {T^2}_2={T^3}_3=-p_t,\, {T^4}_4=\rho
 \end{equation}
 \begin{equation}
 {E^1}_1=-{E^2}_2=-{E^3}_3={E^4}_4=\frac{1}{8\,\pi}\,e^{\nu+\lambda}\,F^{14}\,F^{41}.\\
\end{equation}

Since we are employing spherical symmetry, the
four-current component is only a function of radial distance, $r$.
The only non vanishing components of electromagnetic field tensor
are $F^{41}$ and $F^{14}$, related by $F^{41} = - F^{14}$, which
describes the radial component of the electric field.
If $q(r)$ represents the total charge contained within the
sphere of radius $r$, then it can be defined by the relativistic Gauss law as
\begin{equation}
q(r) = 4\pi \int_0^r \sigma r^2 e^{\lambda/2} dr = r^2 \sqrt{-F_{14}F^{14}}. \label{charge}
\end{equation}

From Eq.(\ref{charge}), we obtain
\begin{equation}
F^{41}=- e^{-(\nu+\lambda)/2}\,\frac{q(r)}{r^2}.
\end{equation}

For the spherically symmetric metric (\ref{metric1}), the
Einstein-Maxwell field equations may be expressed as the following
system of ordinary differential equations \cite{Dionysiou}
\begin{eqnarray}
\hspace{-1.7cm}\frac{{\nu}^{\prime}}{r} e^{-\lambda} - \frac{(1 - e^{-\lambda})}{r^2} = 8\pi p_r - \frac{q^2}{r^4}= -8\pi ({T^1}_1+{E^1}_1) \label{e11}
\end{eqnarray}
\begin{eqnarray}
\left[\frac{{\nu}^{\prime\prime}}{2} - \frac{{\lambda}^{\prime}{\nu}^{\prime}}{4} + \frac{{{\nu}^{\prime}}^2}{4} + \frac{{\nu}^{\prime} - {\lambda}^{\prime}}{2r}\right]e^{-\lambda}= 8\pi p_t + \frac{q^2}{r^4} = -8\pi ({T^2_2}+{E^2}_2) \nonumber\\= -8\pi ({T^3}_3+{E^3}_3)  ,\label{e21}
\end{eqnarray}
\begin{eqnarray}
\hspace{-2.3cm} \frac{{\lambda}^{\prime}}{r} e^{-\lambda} + \frac{(1 - e^{-\lambda})}{r^2}=8\pi \rho + \frac{q^2}{r^4} =8\pi ({T^4}_4+{E^4}_4)  \label{e31}
\end{eqnarray}

where the prime denotes differentiation with respect to $r$. If we define the anisotropy parameter as $\Delta = p_t - p_r$, then from equations (\ref{e11}) and (\ref{e21}) we obtain  \cite{Herrera1985}

\begin{equation} \label{anio}
\Delta = \left[\frac{{\nu}^{\prime\prime}}{2} - \frac{{\lambda}^{\prime}{\nu}^{\prime}}{4} + \frac{{{\nu}^{\prime}}^2}{4}  - \left(\frac{{\nu^\prime} + {{\lambda}^{\prime}}}{2r}\right)\right]e^{-\lambda} + \left(\frac{1 - e^{-\lambda}}{r^2}\right) - 2\frac{q^2}{r^4}
\end{equation}

We note that when $\Delta = 0$ the pressure is isotropic at each interior point of the matter distribution. The term $2(p_t - p_r)/r$
appears in the conservation equations ${T^i}_{j;i} = 0$ and represents
a force due to the anisotropic nature of the fluid. When $p_t > p_r$ the force associated with anisotropy is directed outwards and when  $p_t < p_r$ ($\Delta < 0$), the force is directed inwards. For $\Delta > 0$ it has been shown that it is possible to construct more compact objects compared to their neutral counterparts ($\Delta = 0$)\cite{Gokhroo1994}. Various approaches have been adopted in finding solutions of the above system of equations. The system (\ref{e11}) - (\ref{e31}) consists of six unknowns $\rho$, $p_r$, $p_t$, $\nu$, $\lambda$ and $E^2 = q^2/r^4$, the electric field intensity. In attempting to find exact solutions of the system describing anisotropic charged compact objects one could specify an equation of state of the form $p = p(\rho)$, choose the gravitational potentials $\nu$ and $\lambda$ based on physical grounds or prescribe the behaviour of the anisotropy parameter, $\Delta$.

By using Eqs.(\ref{e11})-(\ref{e31}), the expression of pressure gradient in terms of anisotropy, metric functions and charge read as
\begin{equation}
\frac{dp_r}{dr} =- \frac{\nu'\,(\lambda'+\nu')}{2\,r\,e^{\lambda}}+\frac{q}{4\pi r^4} \frac{dq}{dr} + \frac{2\Delta}{r},\label{e7}
\end{equation}
where the above Eq.~(\ref{e7}) represents
the charged generalization of the well-known
Tolman-Oppenheimer-Volkoff (TOV) equation of hydrostatic for
anisotropic stellar structure \cite{Tolman1939,Oppenheimer1939}.

If the mass function for electrically charged fluid sphere is
denoted by $m(r)$ we may then write
\begin{equation}
\frac{2m(r)}{r}= 1 - e^{-\lambda(r)}  + \frac{q^2}{r^2}.\label{e4}
\end{equation}

Various studies of anisotropic charged compact objects have specified the behaviour of the mass function to obtain the gravitational potential $\lambda(r)$. This is then fed back into the Einstein-Maxwell field equations to obtain the complete gravitational and thermodynamical behavior of the model.

\section{Class one condition for spherical symmetric metric:}

Let us consider a 5-dimensional flat line element

\begin{eqnarray}
\hspace{-2.2cm} ds^{2}=-\left(dz^1\right)^2-\left(dz^2\right)^2-\left(dz^3\right)^2-\left(dz^4\right)^2+\left(dz^5\right)^2,\label{e1}
\end{eqnarray}

\noindent where we suppose the coordinates $z^1$, $z^2$, $z^3$, $z4$ and $z^5$ assume the following forms:\\

$z^1=r\,sin\theta\,cos\phi$, \,\, $z^2=r\,sin\theta\,sin\phi$, \,\,\, $z^3=r\,cos\theta$,\\

$z^4=\sqrt{K}\,e^{\frac{\nu}{2}}\,cosh{\frac{t}{\sqrt{K}}}$,\,\,\,$z^5=\sqrt{K}\,e^{\frac{\nu}{2}}\,sinh{\frac{t}{\sqrt{K}}}$.\\

\noindent In the above transformations $K$ is a positive constant.
We can then write the differential forms of the above components as
 \begin{equation}
\hspace{1cm} dz^1=dr\,sin\theta\,cos\phi + r\,cos\theta\,cos\phi\,d\theta\,-r\,sin\theta\,sin\phi\,d\phi,\label{eq2a}
\end{equation}

\begin{equation}
\hspace{1cm}dz^2=dr\,sin\theta\,sin\phi + r\,cos\theta\,sin\phi\,d\theta\,+r\,sin\theta\,cos\phi\,d\phi,\label{eq2b}
\end{equation}

\begin{eqnarray}
\hspace{-2.7cm} dz^3=dr\,cos\theta\, - r\,sin\theta\,d\theta,\label{eq2c}
\end{eqnarray}

\begin{equation}
dz^4=\sqrt{K}\,e^{\frac{\nu}{2}}\,\frac{\nu'}{2}\,cosh{\frac{t}{\sqrt{K}}}\,dr + e^{\frac{\nu}{2}}\,sinh{\frac{t}{\sqrt{K}}}\,dt,\label{eq2d}
\end{equation}

\begin{equation}
dz^5=\sqrt{K}\,e^{\frac{\nu}{2}}\,\frac{\nu'}{2}\,sinh{\frac{t}{\sqrt{K}}}\,dr + e^{\frac{\nu}{2}}\,cosh{\frac{t}{\sqrt{K}}}\,dt,   \label{eq2e}
\end{equation}
where prime denotes differentiation with respect to the radial coordinate $r$.

\noindent Substituting the above expressions $ dz^1,~dz^2,~ dz^3,~ dz^4 $ and $dz^5$ into the metric (\ref{e1}), we obtain:
\begin{equation}
	ds^{2}=-\left(\,1+\frac{K\,e^{\nu}}{4}\,{\nu'}^2\,\right)\,dr^{2}-r^{2}\left(d\theta^{2}+\sin^{2}\theta d\phi^{2} \right)+e^{\nu(r)}dt^{2},\label{e2}
\end{equation}

\noindent A direct comparison of metric (\ref{e2}) and metric (\ref{metric1}) yields

\begin{equation}
	e^{\lambda}=\left(\,1+\frac{K\,e^{\nu}}{4}\,{\nu'}^2\,\right),\label{eq4}
\end{equation}

The Eq.(\ref{eq4}) provides the embedding class condition.

\noindent The pressure anisotropy factor, ($p_t-p_r=\Delta$), is readily obtained from Eqs. (\ref{e11}-\ref{e21}) together with relation (\ref{eq4}) as

\begin{equation}
\label{delta1}
\Delta=\frac{\nu'\,e^{-\lambda}}{32\,\pi}\,\left(\frac{\nu'\,e^{\nu}}{2B^{2}r}-1\right)\,
\left(\frac{2}{r}-\frac{\lambda'e^{-\lambda}}{1-e^{-\lambda}}\right)-\frac{2q^2}{r^4}.
\end{equation}

It is clear from (\ref{delta1}) that when $\Delta = 0$, we have $p_t = p_r$ at each interior point of the fluid distribution. At this point we should highlight the fact that the embedding class condition (\ref{eq4}) together with pressure isotropy ($\Delta = 0$) yields only {\em two} exact solutions for uncharged fluids, (i) the interior Schwarzschild solution and (ii) the Kohler-Chao solution\cite{Kohler}. The Kohler-Chao solution cannot be used to model a bounded configuration such as a star since there is no surface at which the radial pressure vanishes. Such a surface would define the boundary of the star. The Schwarzschild interior solution describes the interior gravitational field of a uniform density sphere and suffers various pathologies such as the prediction of superluminal propagation velocities within the fluid as well it being unstable against radial perturbations. In a recent paper Maurya and Govender \cite{sungov} modeled charged compact objects with isotropic pressure via embedding. The pressure isotropy condition becomes a definition for the electric field intensity or the charge distribution. Just as in our approach here the embedding relates the two metric functions $\nu(r)$ and $\lambda(r)$. Recently the modeling of compact objects such as neutron stars, pulsars and strange stars has attracted huge attention amongst researchers. This is mainly due to the fact that a large number of data sets are available in the literature against which the strengths and merits of the various theoretical models can be tested. The role of pressure anisotropy within the stellar core has been highlighted in many of these models.\\

\section{Generalized charged anisotropic solution for compact star}

We can recast Eqs. (\ref{e11}), (\ref{e21}) and (\ref{e31}) in terms of mass function as follows:

\begin{eqnarray}
\hspace{-2.2cm} 8\,\pi p_r = \frac{[\nu^{\prime}(r^2+q^2 - 2 r m) - 2m] }{r^3} +\frac{2q^2}{r^4}.\label{pr1}
\end{eqnarray}
\begin{eqnarray}
\hspace{0.3cm} 8\,\pi\, p_t=\frac{[(q q^{\prime} -r m^{\prime})(2+r{\nu}^{\prime})-m(2r^2{\nu}^{\prime\prime} +r^2{{\nu}^{\prime}}^2 +r{\nu}^{\prime}-2)]}{2r^3} \nonumber\\
- \frac{2q^2{\nu}^{\prime} -(r^2+q^2)(2r{\nu}^{\prime\prime} +r{{\nu}^{\prime}}^2+2{\nu}^{\prime})}{4r^3}
- \frac{2q^2}{r^4}\label{pt1}
\end{eqnarray}
\begin{eqnarray}
\hspace{-5.8cm} 8\,\pi \rho= \frac{2m^{\prime}}{r^2} - \frac{2qq^{\prime}}{r^3}\label{d1}
\end{eqnarray}

\noindent In this paper we would like to construct a generalized model by adopting a single generic function $\nu(r)$. The invariance of the   Ricci tensor requires that the energy density $\rho(r)$, radial pressure $p_r(r)$ and
tangential pressure $p_t(r)$ should be finite at the origin. The regularity of Weyl invariants require that mass $m(r)$ and electric charge $q(r)$ should attain minimum values at the centre $r=0$ of the configuration ($m(0)= q(0)=0$)and attain maximum values at the surface of the star. i.e. $m(0)=0$, $m^{\prime}(r) > 0$ and  $q(0)=0$, $ q^{\prime}(r) > 0$.
\noindent In the modeling of charged anisotropic compact stars Maurya et al. \cite{Maurya1} have shown that the metric function $\nu(0)=$ is finite constant, $q(0)= 0$, $\nu^{\prime}(0)=0$ and $\nu^{\prime\prime}(0)> 0$. Since energy density and radial pressure are positive finite and continuous it follows that $r > 2m(r)$ \cite{Baumgarte1993,Mars1996}. Form $p_r \ge 0$  with $r > 2m(r)$ we have $\nu^{\prime}(r)\neq 0$. This shows that the generic function $\nu(r)$ is regular minimum at the centre and monotone increasing function of $r$.
Bearing in mind these observations we suppose the generic function $\nu(r)$ has the following form:

\begin{eqnarray} \label{n10}
\hspace{-1cm}\nu(r)= n\,\ln (1+A\,r^2)+\ln B
\end{eqnarray}

 where we have two cases: \\

 $\hspace{2.5cm}$ Case (i) $n<0$ and $A<0$,\\

$\hspace{2.5cm}$  Case (ii) $n>0$ and $A>0$. \\

\noindent Here $B$ is positive constant. We observe that $\nu(0)=\ln B$, $\nu'=\frac{2\,n\,Ar}{(1+Ar^2)}$ and $\nu''=\frac{2\,nA (1-Ar^2)}{(1+Ar^2)^2}$. It follows that $\nu(0)>0$, $\nu'(0)=0$, $\nu''(0)=2\,nA >0$ and $\nu(r)\neq 0$ with $r\neq 0$ for both cases (i) and (ii). This implies that this generic source function $\nu(r)$ is monotone increasing function of $r$ with regular minimum at $r=0$ (Fig.1). Substituting the value of $\nu$ into Eq.(\ref{eq4}) we obtain

 \begin{equation}
\lambda= \ln[1+C\,Ar^2 (1+Ar^2)^{(n-2)}],\label{eq10}
\end{equation}

\noindent  where $C=n^2\,A\,B\,K$.

This form of the metric function is well-motivated and has been utilised by numerous authors to model compact stars arising from the Karmarkar condition. The parameter $n$ plays a pivot role in the structure and stability of the compact object. Table 5 provides an overview of the class one solutions using the ansatz (\ref{n10}) for the metric function $\nu(r)$. We observe that there is a strong connection the range of $n$ which admits physically viable models and nature of the matter content of the star. Anisotropy and electric charge or the absence thereof dictates the admissibility of the range of $n$. It is evident from Eqn. (\ref{n10}) that in the case of vanishing $n$ the spacetime is rendered flat. In this study we will consider solutions for both $n > 0$ and $n < 0$. This approach will allow us to investigate the impact of the 'switch', $n$ on the various thermodynamical properties of the model.  The solution is not well behaved in range $ -3 < n < 2.7$. For $-7.5 < n \le -3 $ and $2.7 \le n <4$. we will get star with low mass. For $ n \le -7.5 $ and $ n\ge 4 $. we will get stellar models describing compact objects such as Her X-1.

\begin{figure}[h]
\centering
\includegraphics[width=5.5cm]{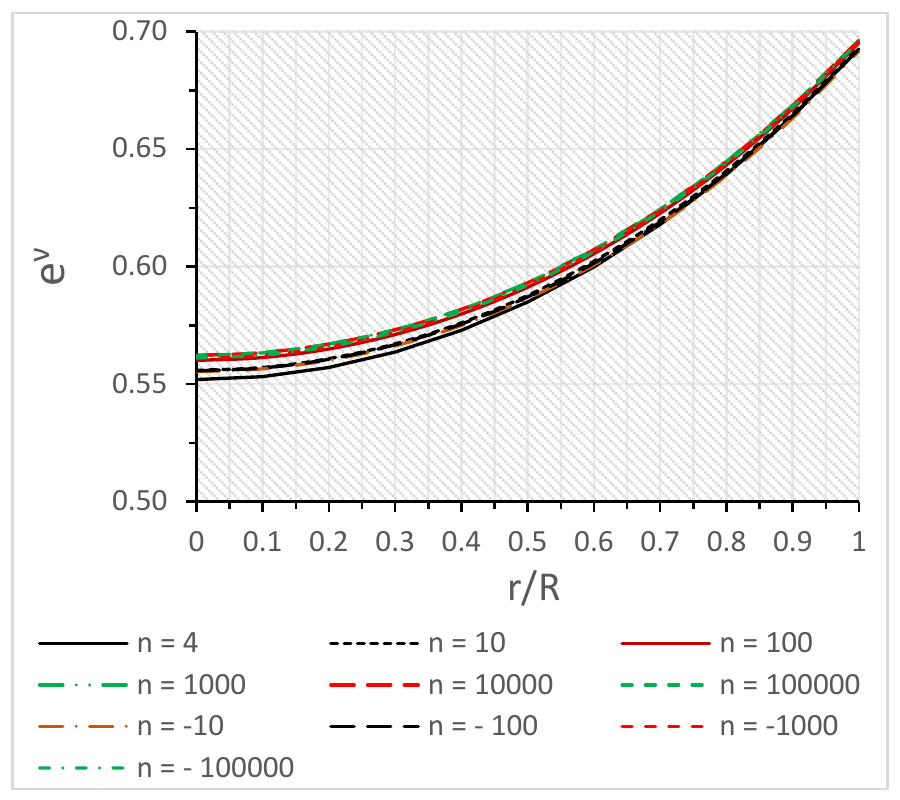} \includegraphics[width=5.5cm]{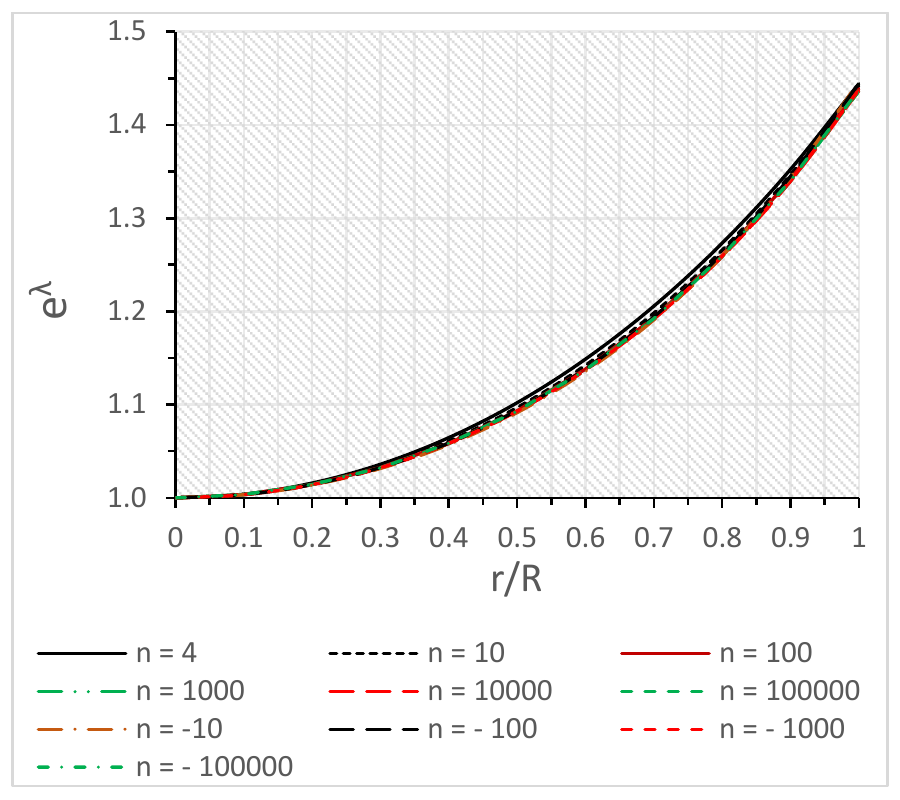}
\caption{Behavior of gravitational potential $e^{\nu}$ (left panel) and $e^{\lambda}$  (right panel) verses fractional radius $r/R$ for Her X-1 . The numerical values of the constants are given in tables 1 $\&$ 2.}
\end{figure}

In order to determine the mass function for electrically charged compact star we suppose the electric charge function $q(r)$ is of the form

\begin{equation}
q(r)= E_0\,A^2\,r^6\,(1 + Ar^2)^n \label{q1}
\end{equation}

\noindent where $E_0$ is positive constant. We note that the electric field $(E = q/r^2)$ vanishes at the center of the configuration. We note that in the case of isotropic pressure the condition of pressure isotropy can be treated as a definition for the charge as a function of the radial coordinate. This approach does not guarantee that $q(r)$ will have physical desirable properties such as the function defined in (\ref{q1}). Utilising Eq.(\ref{e4}) and (\ref{q1}) we readily obtain the mass function $m(r)$ as

\begin{equation}
m(r)= \frac{Ar^3\,(1+Ar^2)^{n-2}\left[C+E_0 Ar^2(1+Ar^2)^2+C E_0\,A^2r^4(1+Ar^2)^{n}\right]}{2} \label{m1}
\end{equation}

 \noindent We observe from Eqs.(\ref{q1}) and (\ref{m1}) that $q(0)=0$ and $m(0)=0$. However both $q'(r)$ and $m'(r)$ are positive for $r>0$ in both cases (i) and (ii). This indicates that $q(r)$ and $m(r)$ are increasing monotonically away from centre and attains regular minimum at $r=0$.

 \begin{figure}[h]
\centering
\includegraphics[width=5.5cm]{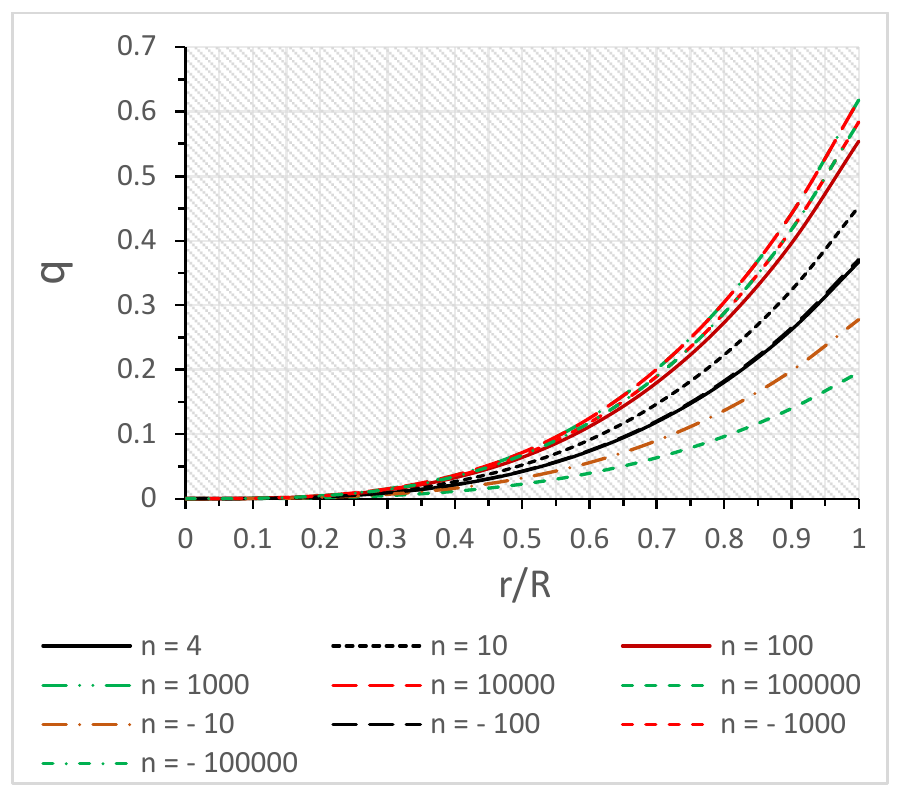} \includegraphics[width=5.5cm]{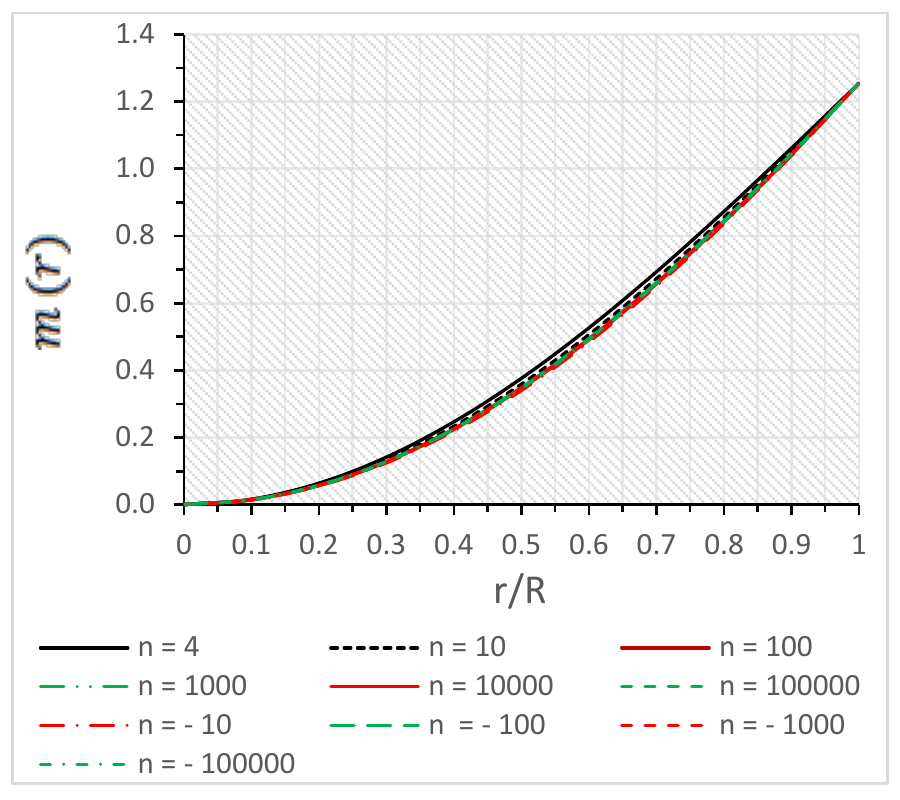}
\caption{Behavior of electric charge $q$  (left panel) and  $m(r)$ versus fractional radius $r/R$ for Her X-1 . The numerical values of the constants are given in tables 1 $\&$ 2.}
\end{figure}

\noindent By plugging the values of the Eqs(\ref{m1}) and (\ref{q1}) into Eqs. (\ref{pr1}), (\ref{pt1}) and (\ref{d1})  we get [by assuming $f=(1+Ar^2)$, $\psi=Ar^2$),
\begin{eqnarray}
\hspace{-3.8cm} p_r=\frac{A\,[2\,n\,+E_0\,\psi\,f^{n+2}+C\,f^n\,(-1+E_0\,\psi^2\,f^n)]}{8\,\pi\,[f^2+C\,\psi\,f^n]};
\end{eqnarray}

\begin{equation}
p_t=\frac{(2+n \psi) n A f^2-A f^n [E_0 (\psi\,f^4+C^2 \psi^3 f^{2 n})+C f (1-\psi+2 E_0 \psi^2 f^{n+1})]}{8\,\pi\,[f^2+C\,\psi\,f^n]^2}
\end{equation}

\begin{equation}
\hspace{-2cm} \rho=\frac{A\,f^n}{8\,\pi}\,\left[\,-E_0\,\psi+\frac{C}{[f^2+C\,\psi\,f^n]}+\frac{2\,C\,f\,[1+(n-1)\,\psi]}{[f^2+C\,\psi\,f^n]^2} \,\right]
\end{equation}

 \begin{figure}[h]
\centering
\includegraphics[width=5.5cm]{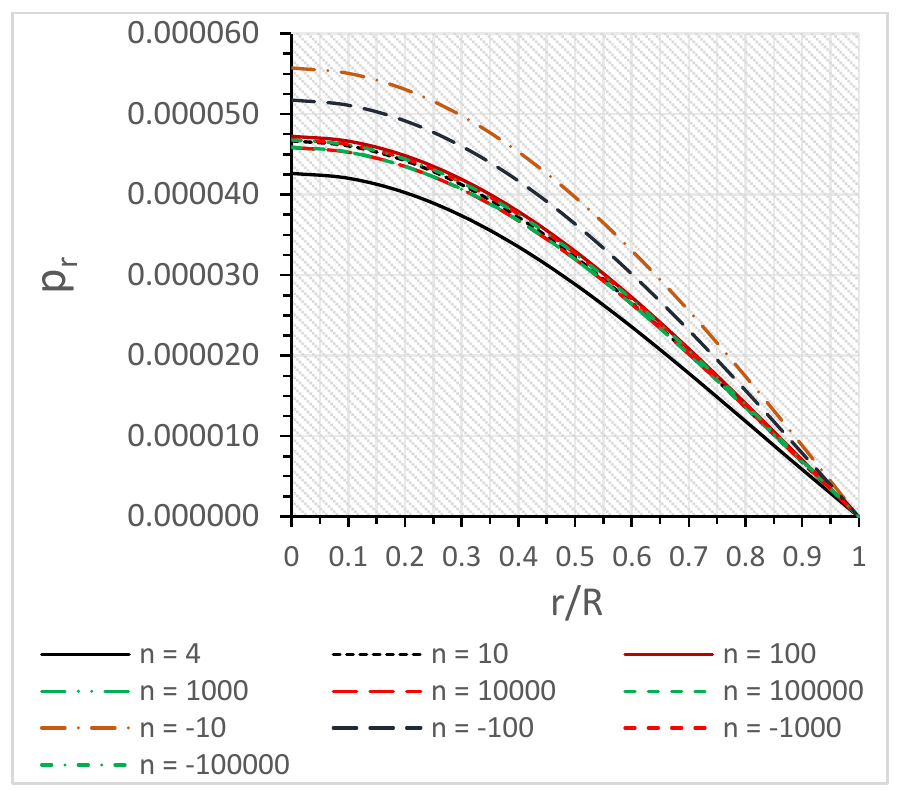} \includegraphics[width=5.5cm]{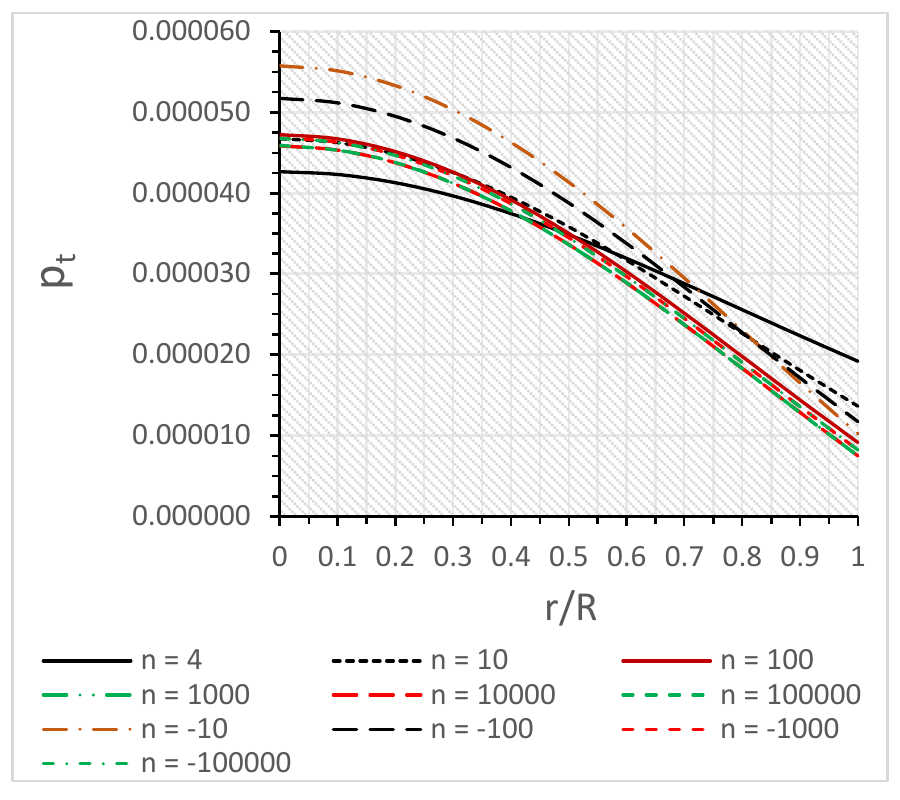}
\caption{Behavior of radial pressure $p_r$  (left panel) and  $p_t$ (right panel) versus fractional radius $r/R$ for Her X-1 . The numerical values of the constants are given in tables 1 $\&$ 2.}
\end{figure}

 \begin{figure}[h]
\centering
\includegraphics[width=5.5cm]{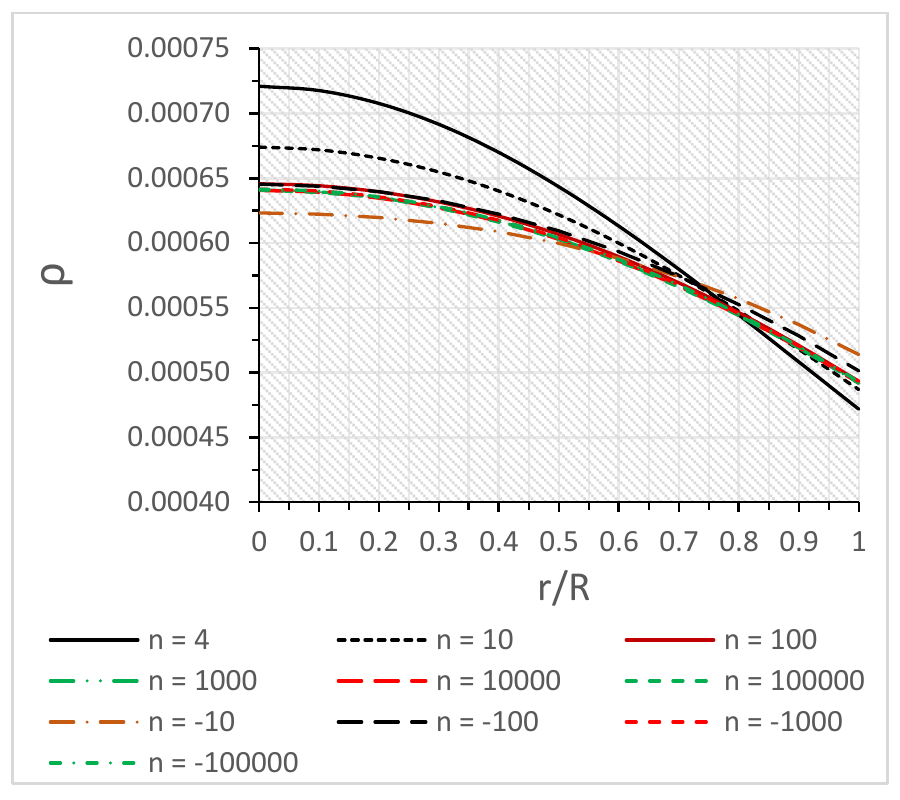}
\caption{Behavior of matter density $ \rho$  versus fractional radius $r/R$ for Her X-1 .The numerical values of the constants are given in tables 1 $\&$ 2. }
\end{figure}

\noindent Relation Eq.(\ref{anio}) provides the anisotropic factor $\Delta$ which is given by

\begin{equation}
\Delta=\frac{\psi\,[-n^2\,f^2+2\,n\,f\,(1+\psi+C\,f^n)+f^n\,(2\,E_0\,f^4+C^2\,f^n\,\Delta_1+2\,C\,f\,\Delta_2)]}{8\,\pi\,[f^2+C\,\psi\,f^n]^2}
\end{equation}

where, ~~~~ $\Delta_1=[-1+2\,E_0\,\psi^2\,f^n]$,\,\, $\Delta_2=[-1+2\,E_0\,\psi^2\,f^{n+1}]$.\\

\begin{figure}[h]
\centering
\includegraphics[width=5.5cm]{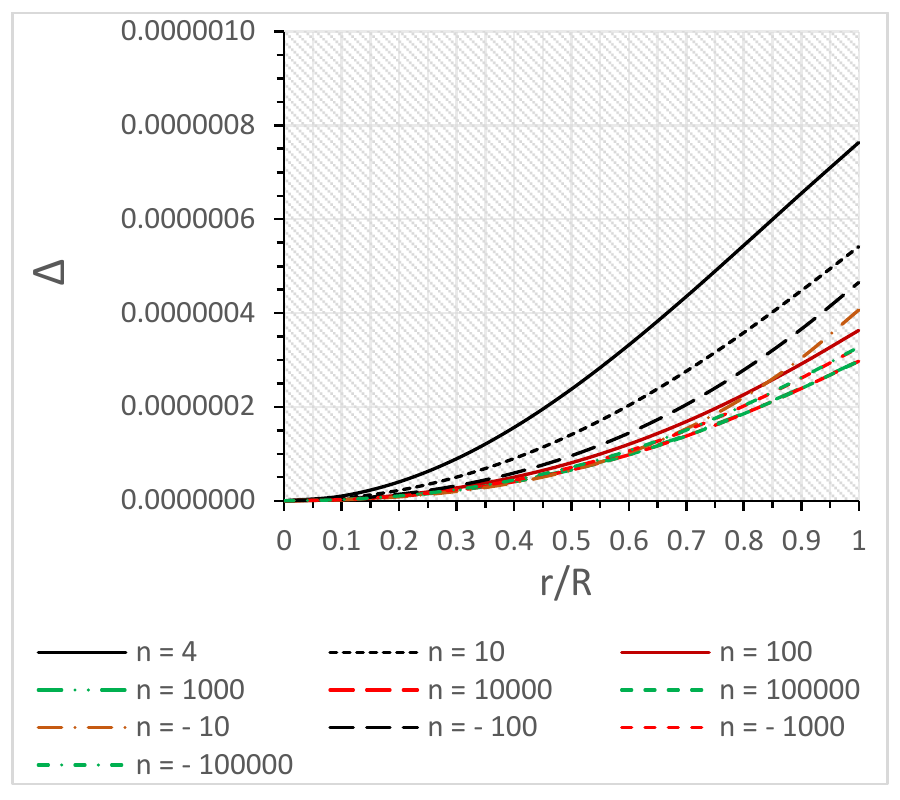}
\caption{Behavior of anisotropic factor $ \Delta$  versus fractional radius $r/R$ for Her X-1 . The numerical values of the constants are given in tables 1 $\&$ 2.}
\end{figure}

\noindent The gradients of the $p_r$, $p_t$ and $\rho$ assume the following forms

\begin{eqnarray}
\hspace{-0.2cm} \frac{dp_r}{dr}=2rA^2\left[\frac{(f^2+C\psi f^n)[Dp_{r1}+n\,f^{n-1}(E_0 \psi f^2-C+C E_0\psi^2 f^n)]+Dp_{r2}\, }{8\,\pi\,[f^2+C\,\psi\,f^n]^2}\right]
\end{eqnarray}

\begin{eqnarray}
\hspace{-1.5cm} \frac{dp_t}{dr}=-2rA^2\left[\frac{Dp_{t1}\,Dp_{t2}+[f+C\, \psi\, f^{n-1}]\,[Dp_{t3}+Dp_{t4}]+f^{n+1} Dp_{t5}}{8\,\pi\,[f^2+C\,\psi\,f^n]^3}\right]
\end{eqnarray}

\begin{eqnarray}
\hspace{-1.5cm} \frac{d\rho}{dr}=2rA^2 f\,\left[\,-E_0\,(f+n\,\psi)\,f^{-1}+\frac{D\rho_1}{[f^2+C\,\psi\,f^n]^3}+\frac{D\rho_2+D\rho_3}{[f^2+C\,\psi\,f^n]^2}\,\right]
\end{eqnarray}

where,
$Dp_{r1}=2\,n+E_0\,f^{n-1}\,[1+3\,\psi^3+\psi\,(5+2\,C\,f^n)+7\,\psi^2\,+C\,(2+n)\,\psi^2\,f^n]$,\\

$Dp_{r2}=-[2+C\,f^n+\psi\,(2+C\,n\,f^{n-1})]\,[2\,n\,f+E_0\,\psi\,f^{2+n}+C\,f^n\,(-1+E_0\,\psi^2\,f^n)]$,\\

$Dp_{t1}=2\,[2+C\,f^n+\psi\,(2+C\,n\,f^{n-1})]$,\\

$Dp_{t2}=(2+n\,\psi)\,n\,f^2-f^n\,[E_0\,\psi\,f^4+C^2\,E_0\,\psi^3\,f^{2n}+C\,f\,(1-\psi+2\,E_0\,\psi^2\,f^{n+1}]$,\\

$Dp_{t3}=-n^2f^2(1+3\psi)-n[4-C f^n-E_0\psi^5f^n-4E_0\psi^4 f^n(1+Cf^n)+\psi\,(8-E_0 f^n)]$,\\

$Dp_{t4}=E_0\,\psi^3\,f^n\,(6+8\,C\,f^n+3\,C^2\,f^{2n})-\psi^2\,[4-4\,E_0\,f^n+C\,f^n\,(1-4\,E_0\,f^n)]$,\\

$Dp_{t5}=-2\,C\,\psi+E_0\,[1+5\,\psi^4+4\,\psi\,(2+C\,f^n)(1+2\,\psi)+3\,\psi^2\,(6+4\,C\,f^n+C^2\,f^{2n})]$,\\

$D\rho_1=-4\,C\,f\,[1+(n-1)\psi]\,[2+C\,f^n+\psi\,(2+C\,n\,f^{-1+n})]$,\\

$D\rho_2=2\,C\,(n-1)\,f+2\,C\,[1+(n-1)\psi]-C\,[2+C\,f^n+\psi\,(2+C\,n\,f^{n-1})]$,\\

$D\rho_3=nC\,f^{-1}\,[f^2+C\,\psi\,f^n]+2\,C\,f\,[1+(n-1)\,\psi]$.\\

The physical viability of our model will be pursued in the next section.

\section{Physical properties of the solution}

Since pressure and density must be positive and finite at centre i.e.  $(p_r)_{r=0}>0$,  $(p_t)_{r=0}>0$ and $(\rho)_{r=0}>0$. Also pressure and density must attain maximum at centre and decrease continuously throughout the star i.e. $\left(\frac{d^2p_r}{dr^2}\right)_{r=0}<0$, $\left(\frac{d^2p_t}{dr^2}\right)_{r=0}<0$ and $\left(\frac{d^2\rho}{dr^2}\right)_{r=0}<0$.  For this purpose we have calculated:

\begin{equation}
(p_r)_{r=0}=(p_t)_{r=0}=\frac{A\,(2n-C)}{8\,\pi} > 0  \label{p0}
\end{equation}

\begin{equation}
(\rho)_{r=0}=\frac{3AC}{8\,\pi} >0  \label{d0}
\end{equation}

\begin{equation}
\left(\frac{d^2p_r}{dr^2}\right)_{r=0}=\frac{2A^2\,[C^2+E_0-C\,(3n-2)-2n]}{8\,\pi} <0  \label{dpr0}
\end{equation}

\begin{eqnarray}
\hspace{1.2cm} \left(\frac{d^2p_t}{dr^2}\right)_{r=0}=\frac{2A^2\,[2C^2-E_0-C(5n-4)+n\,(n-4)]}{8\,\pi} <0 \label{dpt0}
\end{eqnarray}

\begin{eqnarray}
\hspace{-0.5cm} \left(\frac{d^2\rho}{dr^2}\right)_{r=0}=\frac{2A^2\,[-5C^2-E_0+5\,C\,(n-2)]}{8\,\pi} <0 \label{dd0}
\end{eqnarray}

\noindent Case(i). if $n<0$ and $A<0$ : Then the Eqs. (\ref{p0}) and (\ref{d0}) provide $C>2n$ and $C<0$.  However the Eqs.(\ref{dpr0}-\ref{dd0}) give $0\le E_0 < n^2$.\\

\noindent Case (ii). if $n>0$ and $A>0$ : The Eqs. (\ref{p0}) and (\ref{d0}) provide $C<2n$ and $C>0$.  However the Eqs.(\ref{dpr0}-\ref{dd0}) give $0\le E_0 < n^2$.\\

\section{Junction conditions}

In order to generate a model of a physically realizable bounded object we need to ensure that the interior spacetime ${\cal M}^{-}$ must match smoothly to the exterior spacetime ${\cal M}^{+}$. Since the exterior spacetime is empty, ${\cal M}^{+}$ is taken to be the Reissner-Nordstrom solution.

The boundary of the star is that surface for which the radial pressure vanishes, $p_r=0$ at $r=R$ (Misner and Sharp [44]). For our model we obtain

\begin{equation}
C=\frac{(1+A\,R^2)^{1-n}\,[2\,n+E_0\,A\,R^2\,(1+A\,R^2)]}{(1-E_0\,A^2\,R^4\,(1+A\,R^2)^n)}
\end{equation}

The constant $B$ can be determined by using the condition $e^{\nu(R)}=e^{-\lambda(R)}$, which yields:

\begin{equation}
B=\frac{1}{(1-AR^2)^n\,[1+C\,AR^2\,(1-AR^2)^{n-2}]}   \label{eq17}
\end{equation}

However the constant $A$ can be determined using the surface density $\rho_s$ of the star.

\subsection{Energy conditions}
The charged anisotropic fluid sphere should satisfy the following three energy conditions, viz., (i) null energy condition (NEC), (ii) weak energy
condition (WEC) and (iii) strong energy condition (SEC).
For satisfying the above energy conditions, the following inequalities must be hold simultaneously inside the charged fluid sphere:

\begin{eqnarray}
\hspace{-2cm} NEC: \rho+\frac{E^2}{8\pi}\geq 0,\label{eq22}
\end{eqnarray}
\begin{eqnarray}
\hspace{0.8cm} WEC: \rho+p_r \geq  0 , ~~~\,\, \rho+p_t + +\frac{E^2}{4\pi} \geq 0 \label{eq23}
\end{eqnarray}
\begin{eqnarray}
\hspace{-1.4cm} SEC: \rho+p_r+2p_t+\frac{E^2}{4\pi} \geq  0.\label{eq24}
\end{eqnarray}

\begin{figure}[h]
\centering
\includegraphics[width=5cm]{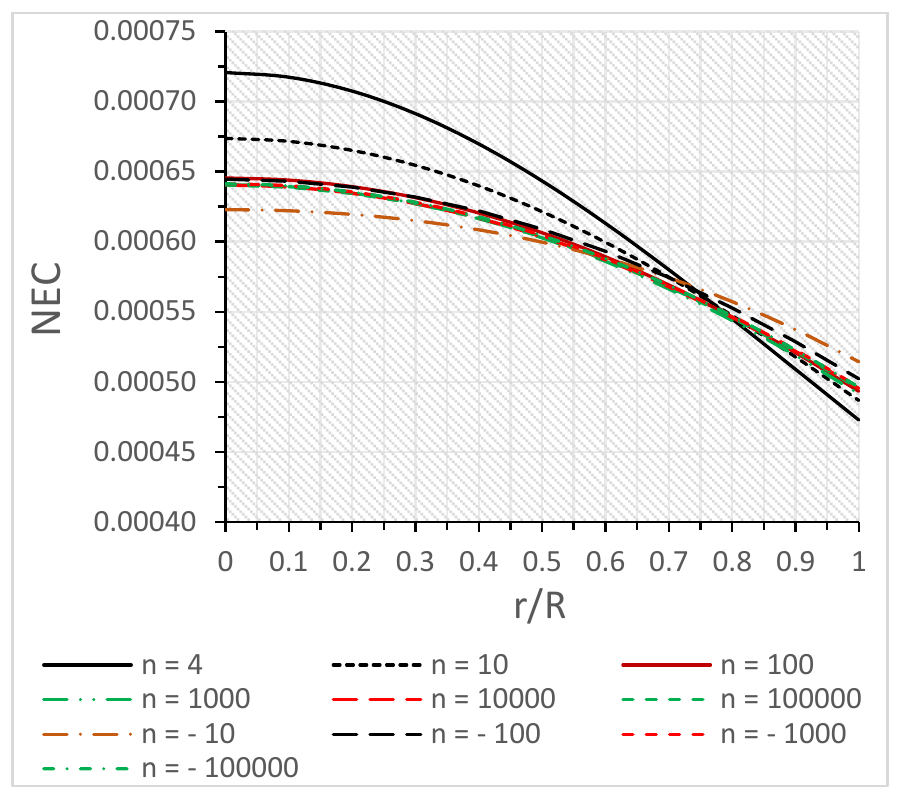} \includegraphics[width=5cm]{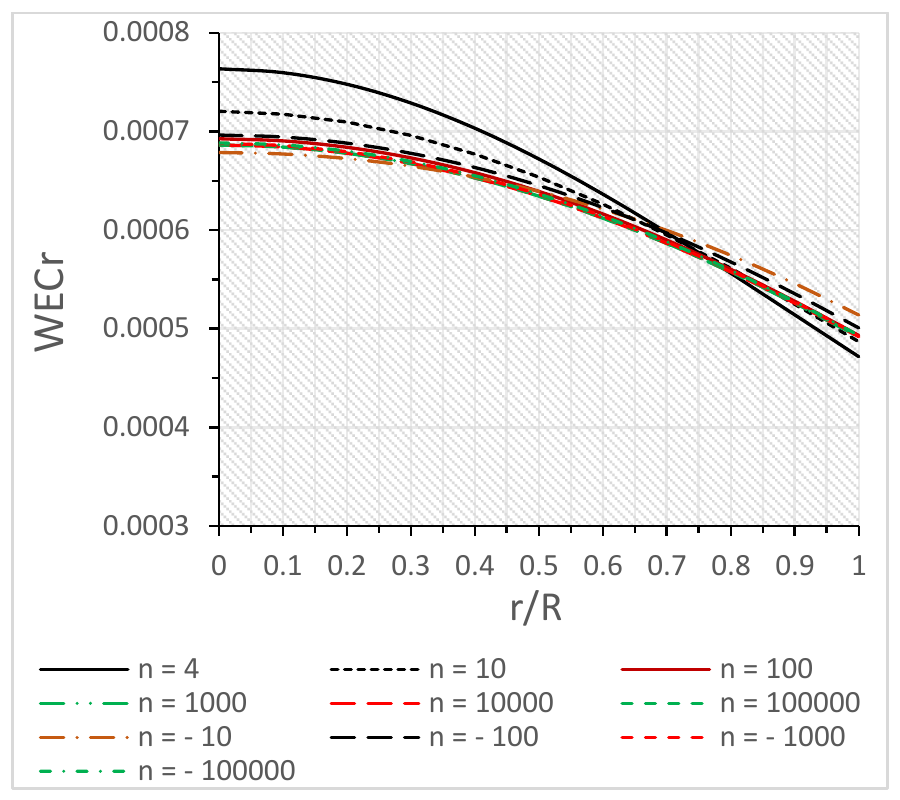}
\includegraphics[width=5cm]{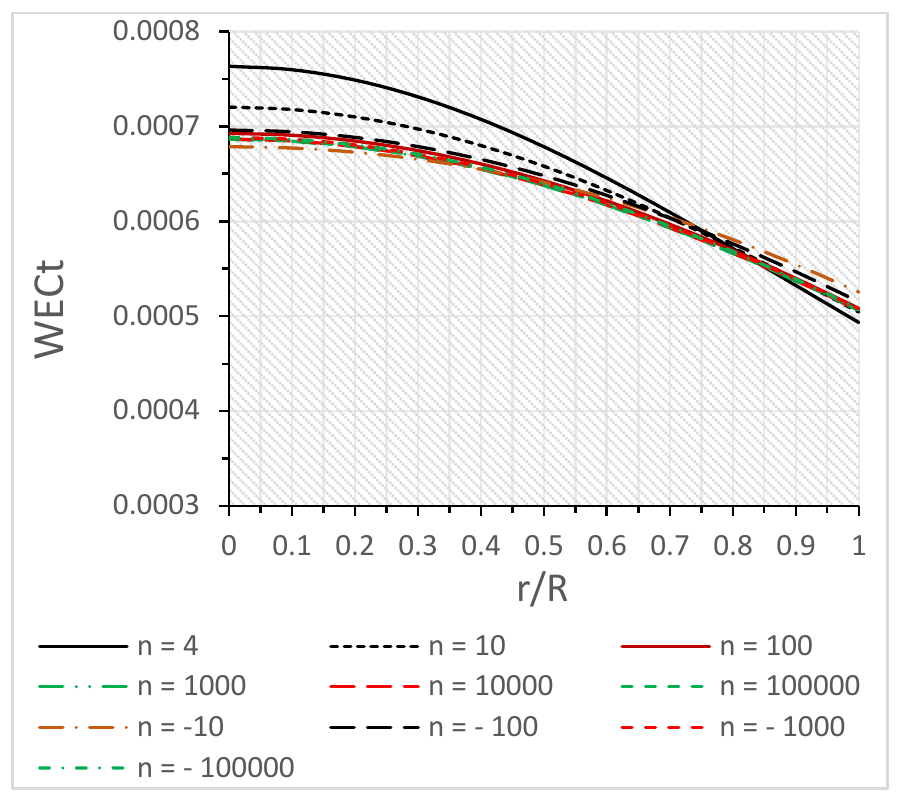} \includegraphics[width=5cm]{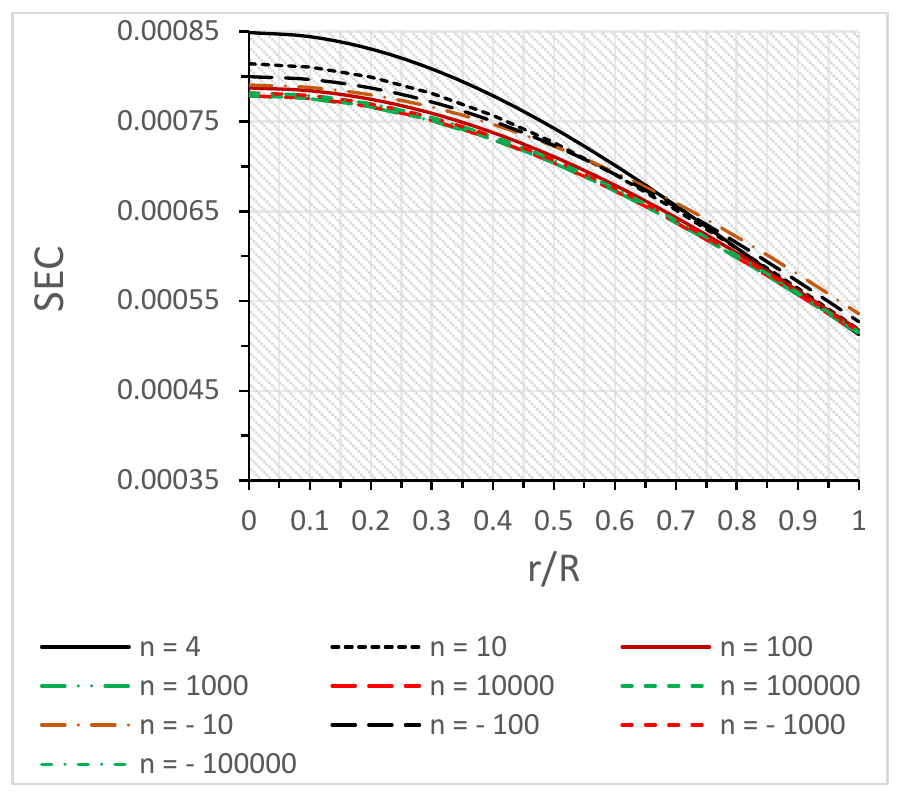}
\caption{Behavior of various energy conditions versus fractional radius $r/R$ for Her X-1 . The numerical values of the constants are given in tables 1 $\&$ 2.}
\label{energy1}
\end{figure}

\subsubsection{Equilibrium condition}

The Tolman-Oppenheimer-Volkoff (TOV) equation~\cite{Tolman1939,Oppenheimer1939} in the presence of charge is given from Eq. (\ref{e7}) as
\begin{equation}
\frac{dp_r}{dr} =- \frac{\nu'\,(\lambda'+\nu')}{2\,r\,e^{\lambda}}+\frac{q}{4\pi r^4} \frac{dq}{dr} + \frac{2\Delta}{r}, \label{eq25}
\end{equation}

\noindent The above equation can be expressed into four different components gravitational force $F_g=- \frac{\nu'\,(\lambda'+\nu')}{2\,r\,e^{\lambda}}$, hydrostatic force $F_h=- dp_r/dr$ , electric force $F_e=\frac{q}{4\,\pi r^4}\,\frac{dq}{dr}$ and anisotropic force $F_a=\frac{2\Delta}{r}$ which are defined as:

\begin{eqnarray}
\label{eq28}
\hspace{2cm} F_g=- \frac{2\,n\,A^2\,r\,\,[C (1 -Ar^2)\,f^n + n\,[f^2 + 2C Ar^2 f^n])}{[f^2 + 2C Ar^2 f^n]^2}
\end{eqnarray}
\begin{eqnarray}
\hspace{-4cm} F_h=-\frac{dp_r}{dr}  \label{eq29}
\end{eqnarray}
\begin{eqnarray}
\hspace{-0.5cm} F_e=\frac{2 r E_0 A^2 f^{n-1}[3+(n+3) Ar^2]}{8\,\pi} \label{eq30}
\end{eqnarray}
\begin{eqnarray}
\hspace{-4.2cm} F_a=\frac{2\Delta}{r}  \label{eq31}
\end{eqnarray}

Observations of the various panels in Fig. 7 show that the force due to anisotropy dominates the electromagnetic force for small $|n|$. As $|n|$ increases the difference in the magnitudes of $F_a$ and $F_e$ decrease until they are equal for a particular value of $|n|$. A further increase in $|n|$ shows that $F_e$ dominates $F_a$ with the relative difference being more marked for large positive values of $n$.

\begin{figure}[h]
\centering
\includegraphics[width=4cm]{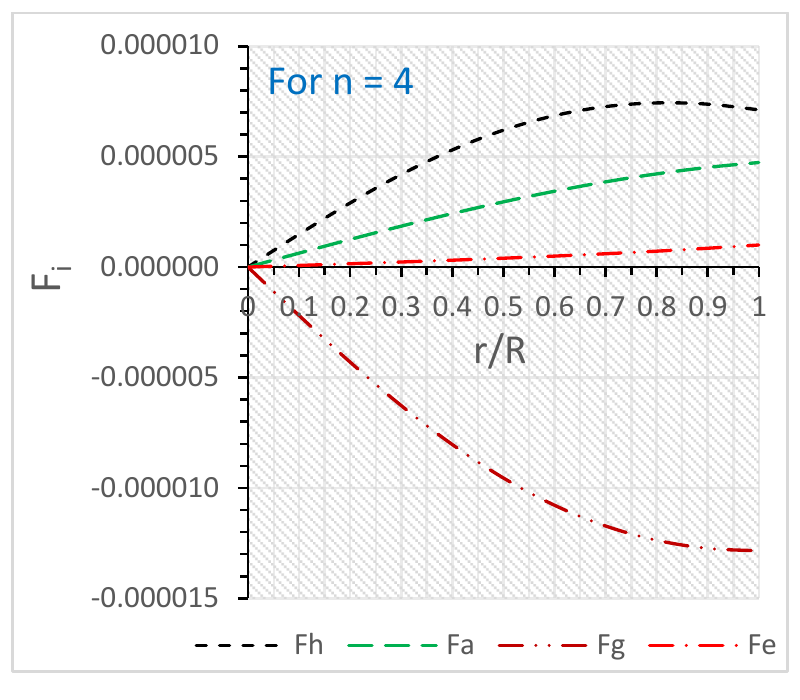} \includegraphics[width=4cm]{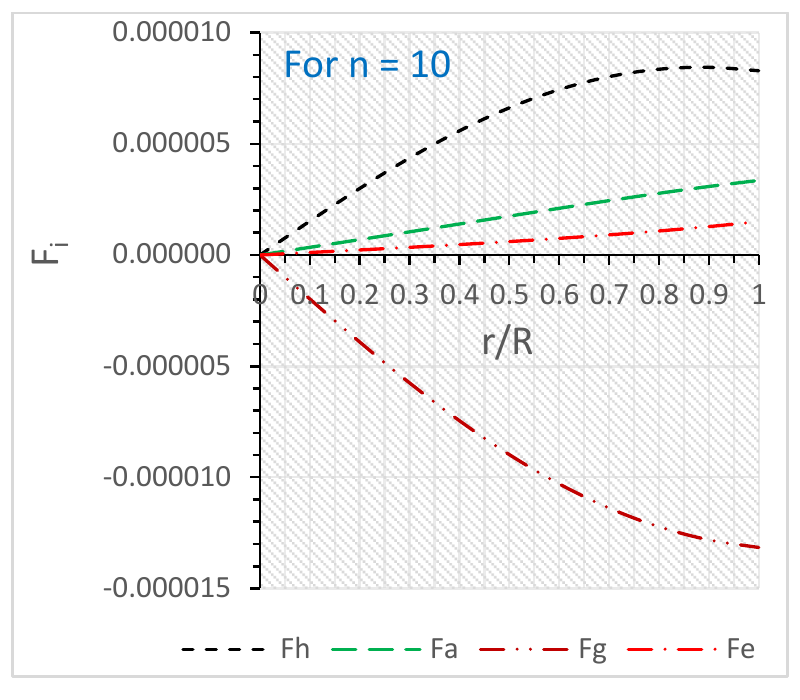}\includegraphics[width=4cm]{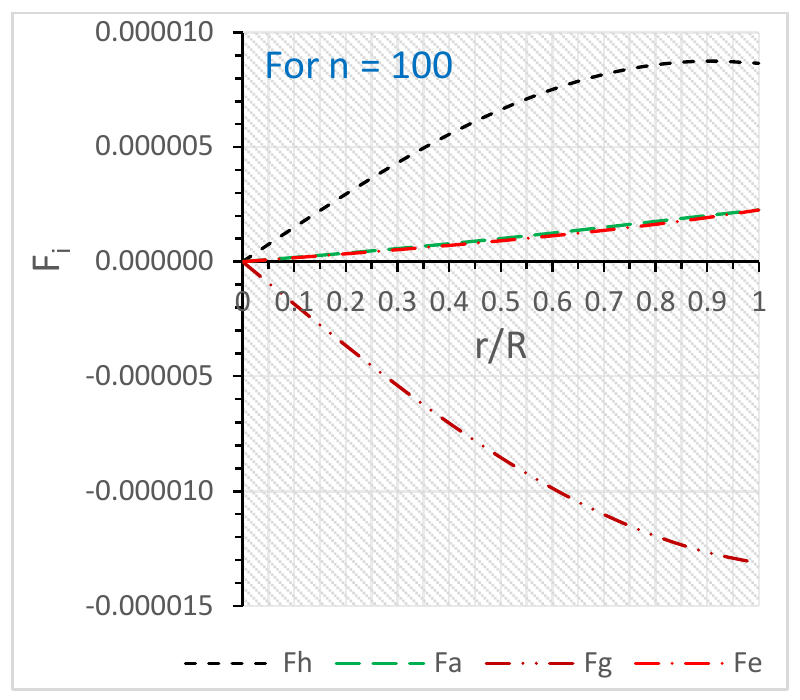}
 \includegraphics[width=4cm]{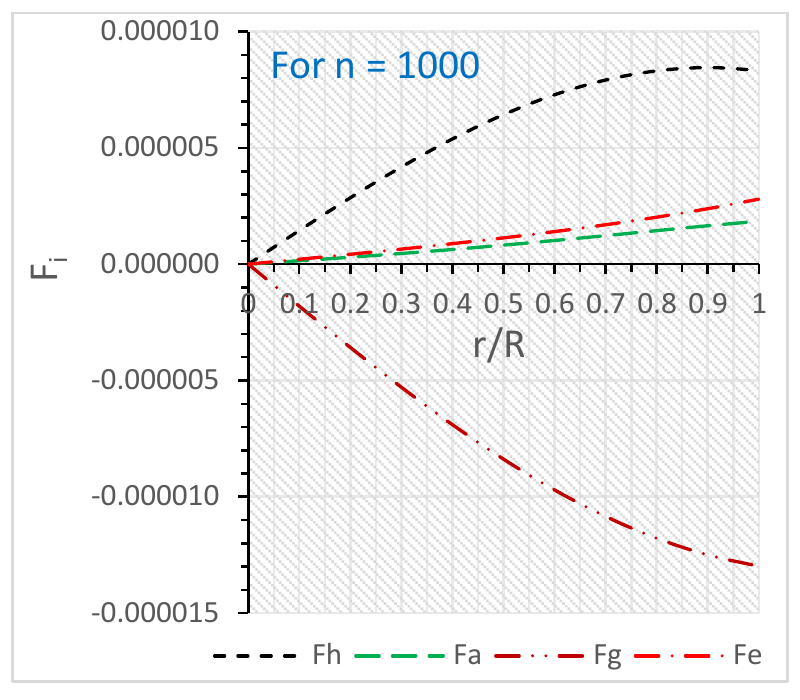} \includegraphics[width=4cm]{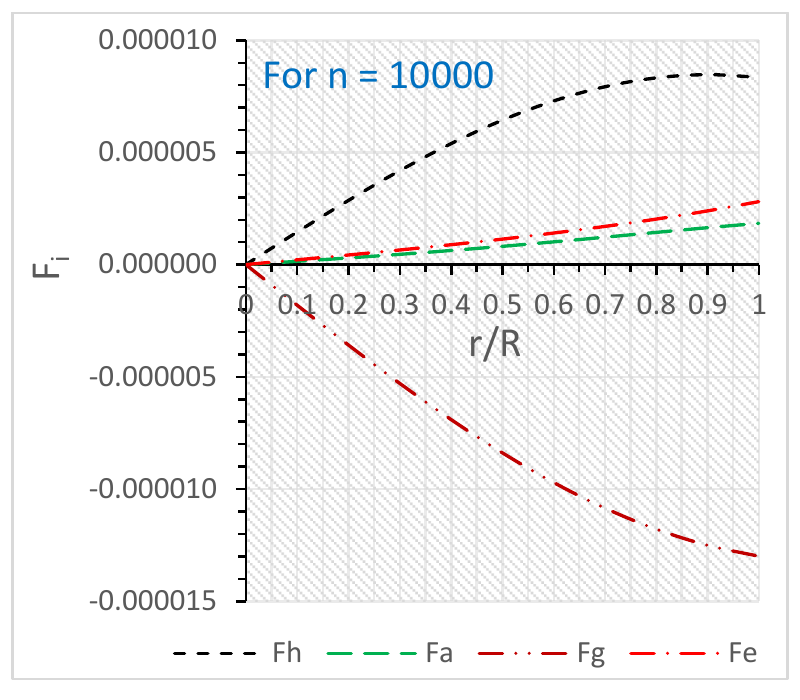} \includegraphics[width=4cm]{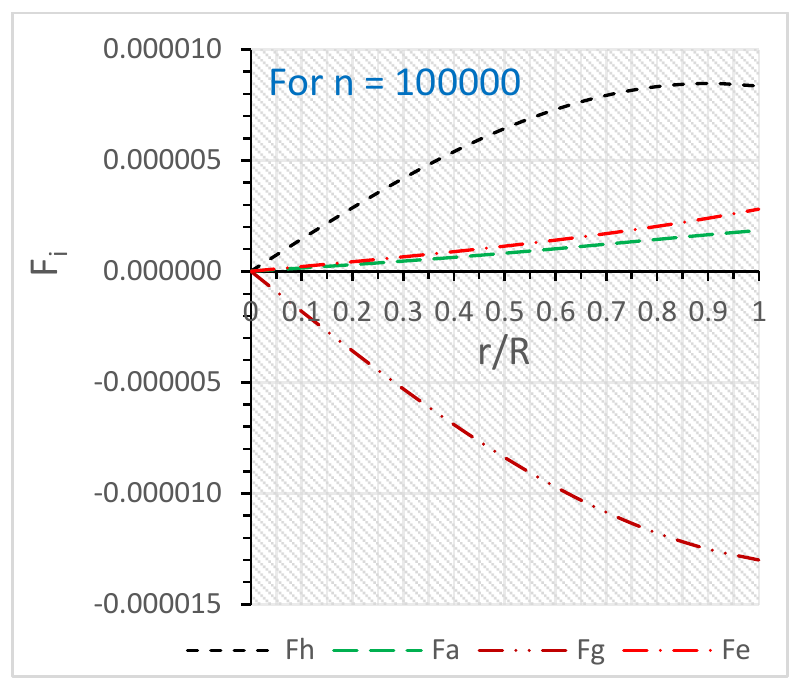}
 \includegraphics[width=4cm]{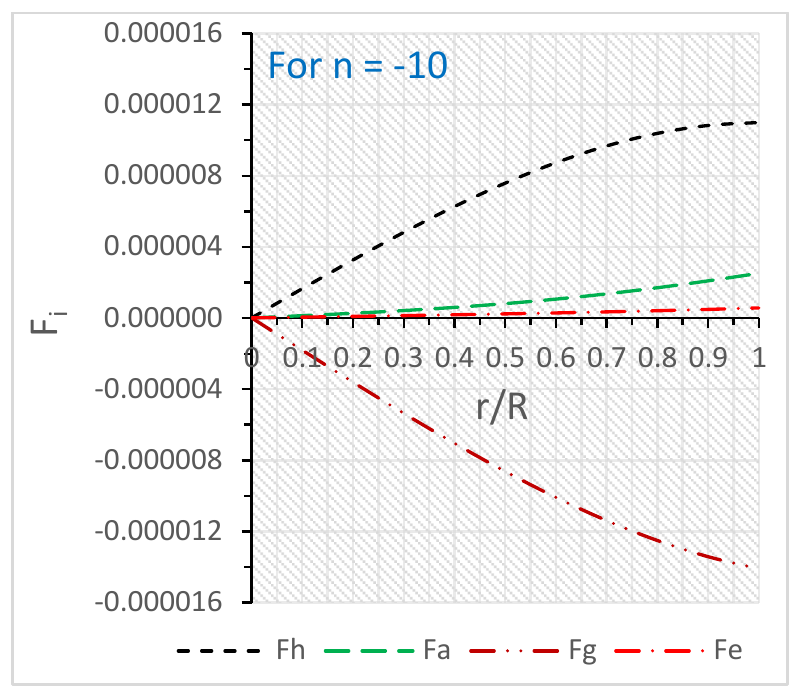} \includegraphics[width=4cm]{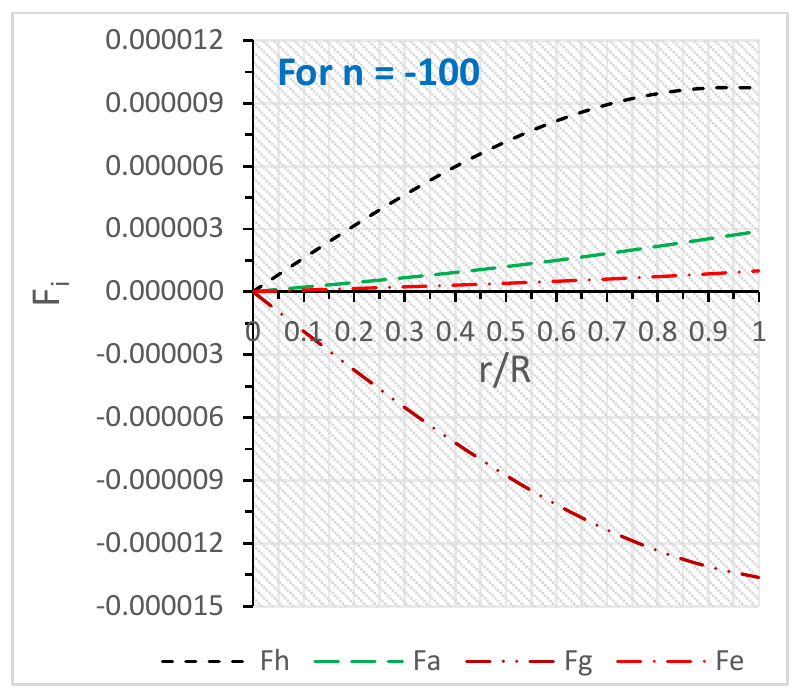} \includegraphics[width=4cm]{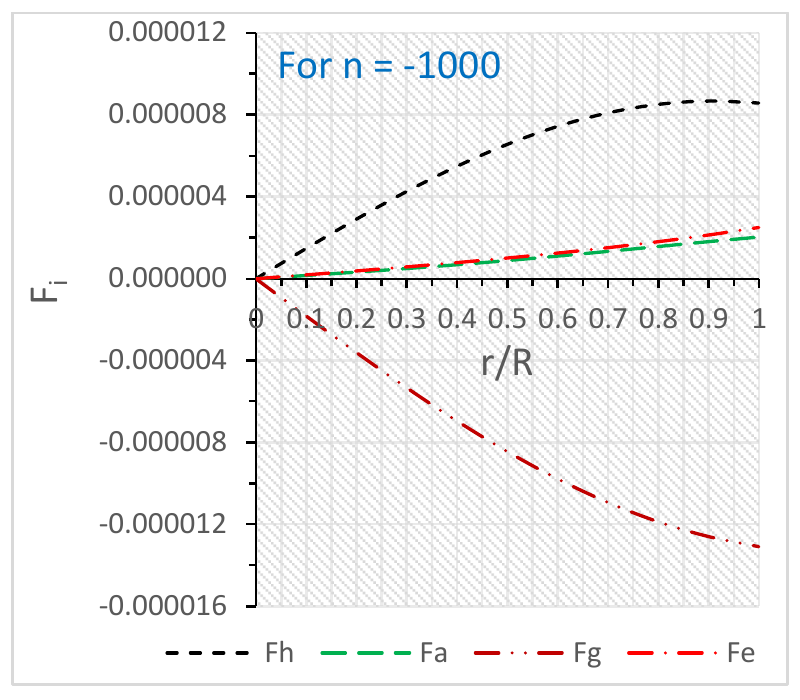}
 \includegraphics[width=4cm]{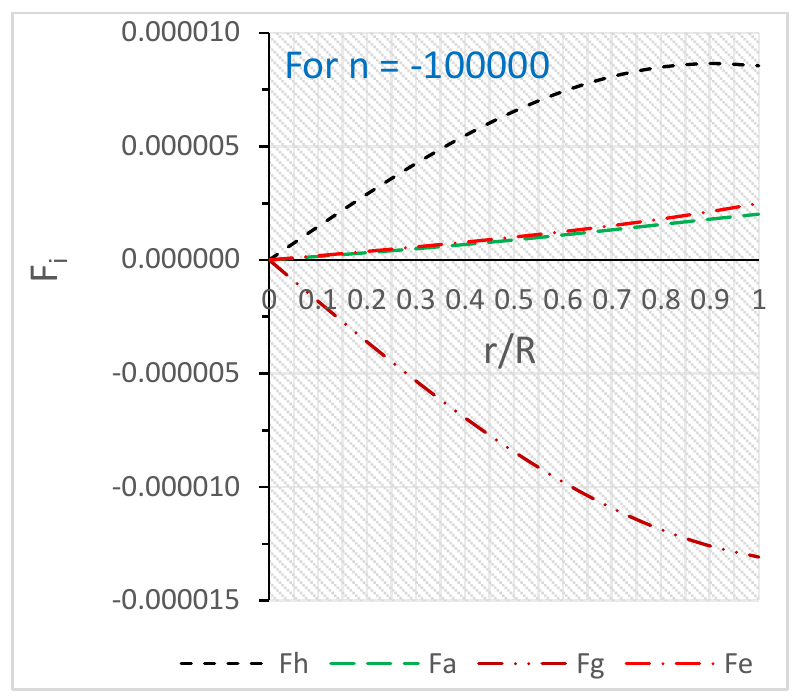}
\caption{Behavior of different forces versus fractional radius $r/R$ for Her X-1. The numerical values of the constants are given in tables 1 $\&$ 2.}
\end{figure}

\subsubsection{velocity of sound \& stability analysis:}

The causality condition should be obeyed i.e. velocity of sound should be less than that of light throughout the model. In addition to the above the velocity of sound should be decreasing towards the surface i.e.${d \over dr}~{dp_r \over d\rho}<0$  or   ${d^2 p_r\over d\rho^2}>0$   and ${d\over dr}{dp_t \over d\rho}<0$  or   ${d^2p_t \over d\rho^2}>0$  for $0\leq r\leq r_b$ i.e. the velocity of sound is increasing with the increase of density and it should be decreasing outwards. From Fig. 8 we observe that the speed of sound decreases monotonically from the center of star (high density region) towards the surface of the star (low density region). The sound speed is less than unity thus indicating that causality is preserved within the stellar core (Fig. 9). In their study of the stability of relativistic spheres, Herrera et al. adopted a perturbative scheme in which the energy density and the anisotropy were perturbed and the effect of these perturbations on the fluid elements were studied. They were able to show that different parts of the star respond differently to various degrees of anisotropy which may lead to cracking or overturning within the core. Abreu et al. utilised a different approach to studying cracking in static spheres. In their approach the difference in the tangential and radial sound speeds served as an indicator of potentially unstable regions. They further showed that stable regions within the stellar fluid are characterised by the stability factor,  $|v_t^2-v_r^2|$ which has to be less than unity for a potentially stable configuration. Fig. 10 clearly indicates that our model is stable for a large range of $|n|$.

\begin{figure}[h]
\centering
\includegraphics[width=5cm]{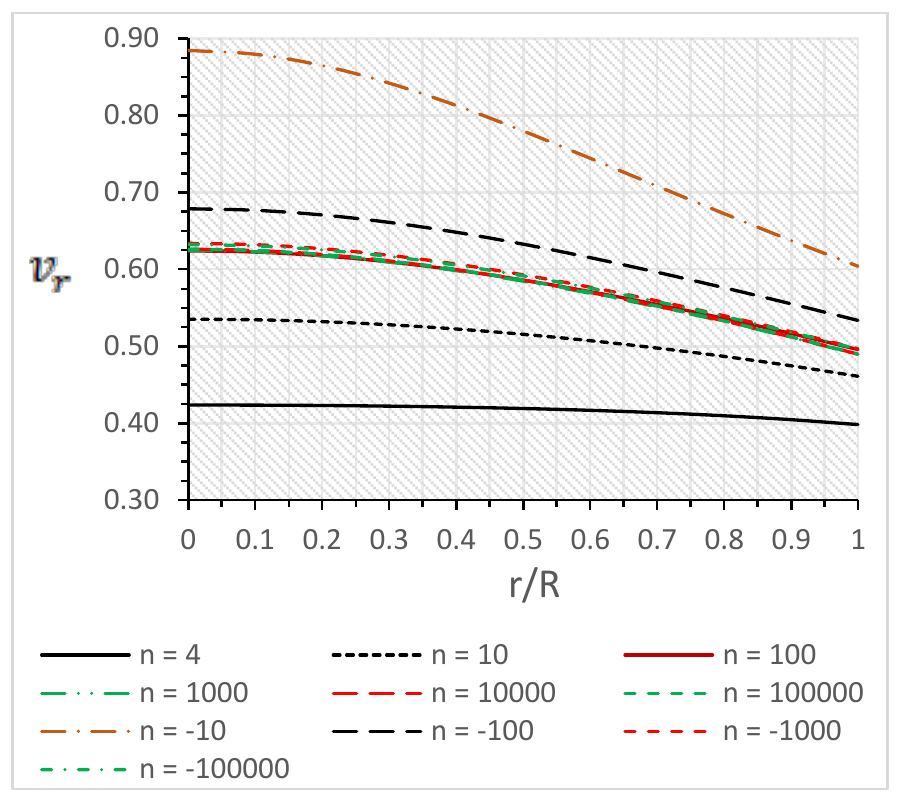} \includegraphics[width=5cm]{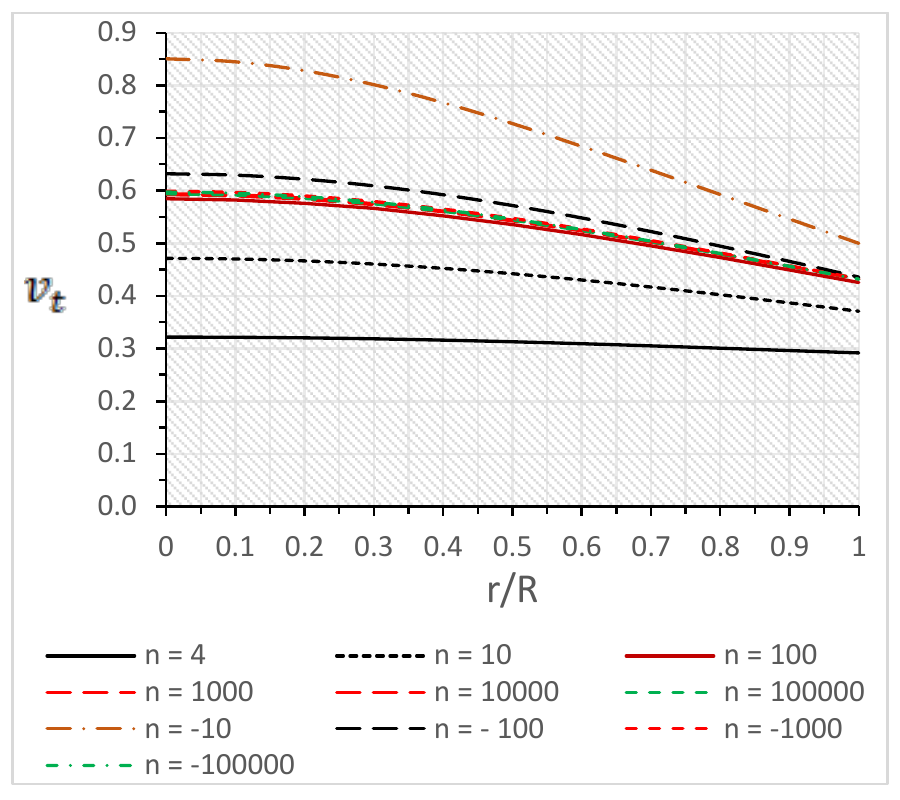}
\caption{Behavior of radial velocity $v_r$ (left panel) and tangential velocity $v_t$ (right panel) verses fractional radius $r/R$ for Her X-1. The numerical values of the constants are given in tables 1 $\&$ 2.}
\label{energy1}
\end{figure}

\begin{figure}[h]
\centering
\includegraphics[width=5cm]{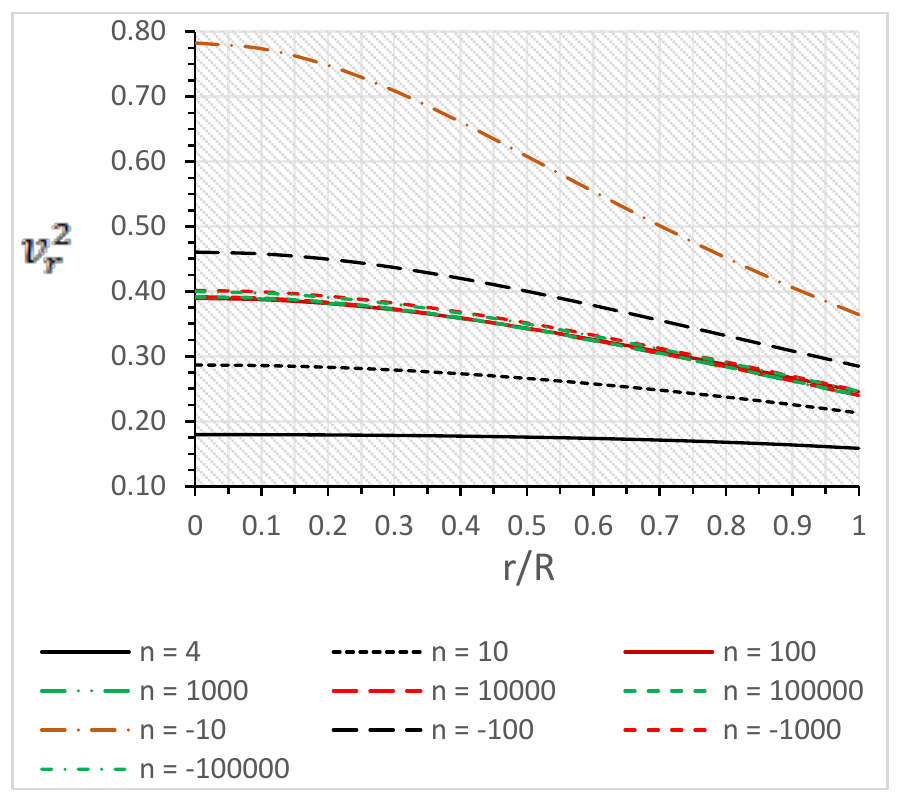} \includegraphics[width=5cm]{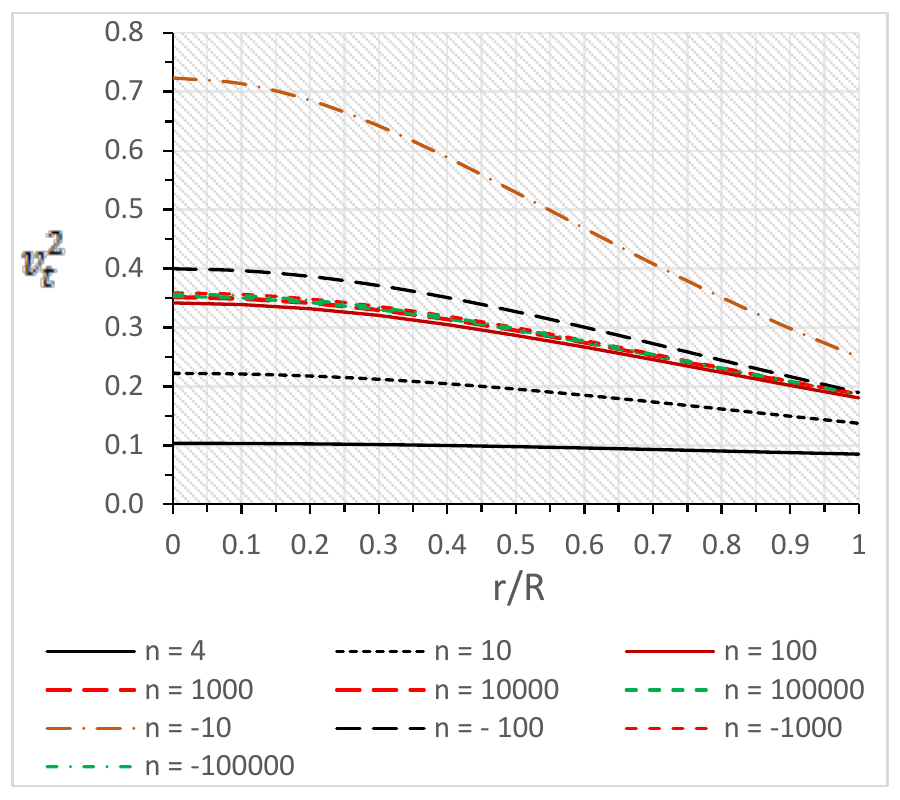}
\caption{Behavior of radial velocity $v^2_r$ (left panel) and tangential velocity $v^2_t$ (right panel) versus fractional radius $r/R$ for Her X-1. The numerical values of the constants are given in tables 1 $\&$ 2.}
\label{energy1}
\end{figure}

\begin{figure}[h]
\centering
\includegraphics[width=5cm]{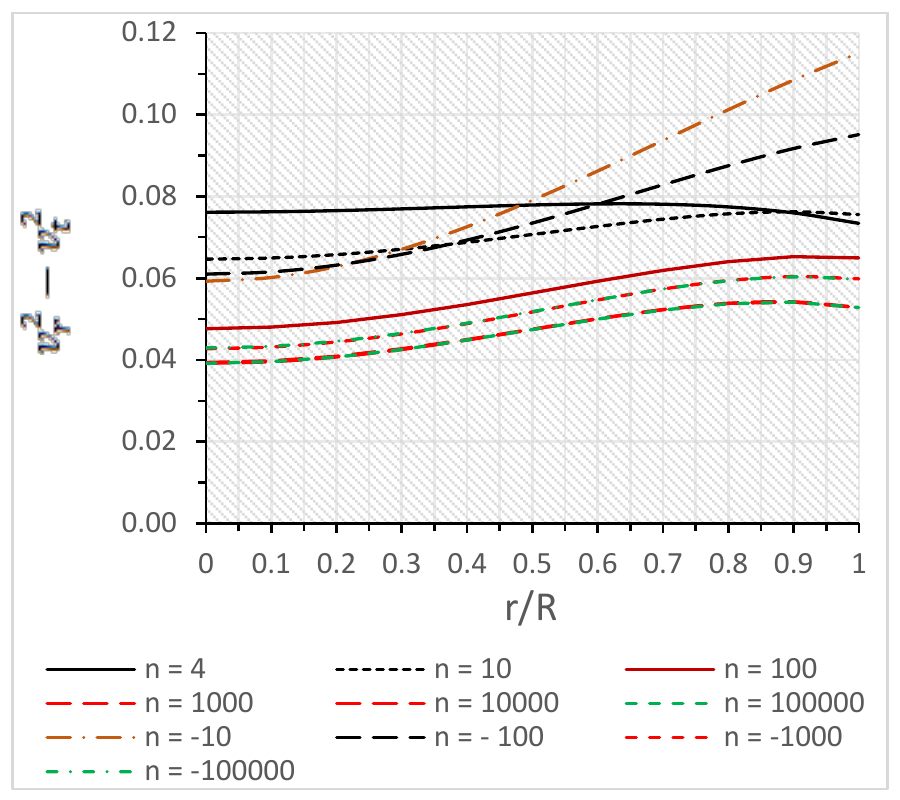} \includegraphics[width=5cm]{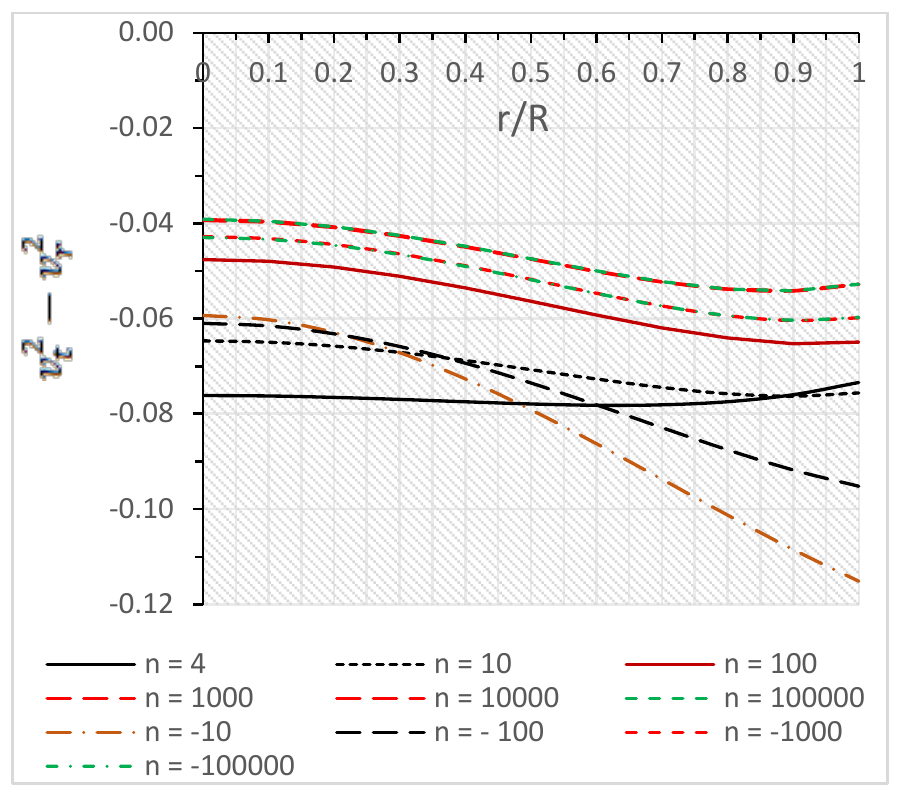}
\caption{Behavior of radial velocity $v^2_r-v^2_t$ (left panel) and tangential velocity $v^2_t-v^2_r$ (right panel) versus fractional radius $r/R$ for Her X-1. The numerical values of the constants are given in tables 1 $\&$ 2.}
\label{energy1}
\end{figure}

\subsubsection{Maximum allowable mass and redshift:}

The well-known Buchdahl\cite{Buchdahl} limit for relativistic static spheres $2M/R\leq 8/9 $, has been generalised for static charged spheres. Work by Andr{\'e}asson~\cite{And} and B{\"o}hmer and Harko~\cite{Boehmer2006} showed that mass to radius ratio in the presence of charge was restricted to
\begin{equation}
\frac{Q^{2}\, (18 R^2+ Q^2) }{2R^{2}\, (12R^2+Q^2)}  \leq
\frac{M}{R} \leq \left[\frac{4R^2+3Q^2}{9R^2} +\frac{2}{9R}\,\sqrt{R^2+3Q^2}\right]
\end{equation}
The compactness $u(r)$ can be defined in terms of the effective mass, $m_{eff}$:

\begin{equation}
u(R)=\frac{m_{eff}(R)}{R}=\frac{1}{2}[1-e^{-\lambda(R)}]\, \label{eq33}
\end{equation}
where
\begin{equation}
m_{eff}=4\pi{\int^R_0{\left(\rho+\frac{E^2}{8\,\pi}\right)\,r^2\,dr}}=\frac{R}{2}[1-e^{-\lambda(R)}]\, \label{eq32}
\end{equation}
and the metric potential $e^{-\lambda}$ is given in (\ref{eq10}).
In their study of anisotropic static spheres, Bowers and Liang\cite{bowers} showed that the surface redshift can be arbitrarily large.
In the case of isotropic stars the surface redshift has an upper bound of $Z_s = 4.77$. The relative magnitude of the radial and tangential stresses within the core plays an important role in determining the magnitude of $Z_s$. As pointed out by Maurya et al. \cite{sungov} when $p_t > p_r$ the associated surface redshift is greater than its isotropic counterpart. The gravitational surface red-shift ($Z_s$) can be calculated from:

\begin{equation}
Z_s= (1-2\,u)^{\frac{-1}{2}} -1= \sqrt{1+C\,Ar^2 (1+Ar^2)^{(n-2)}}-1, \label{zs}
\end{equation}

We note that the surface redshift depends on the compactness $u$ which should in principle, be constrained the Buchdhal limit. Tables 3. and 4. show that the surface redshift decreases with an increase in $|n|$ (Fig. 11). For very large $|n|$ the surface redshift is approximately constant.

\begin{figure}[h]
\centering
\includegraphics[width=5.5cm]{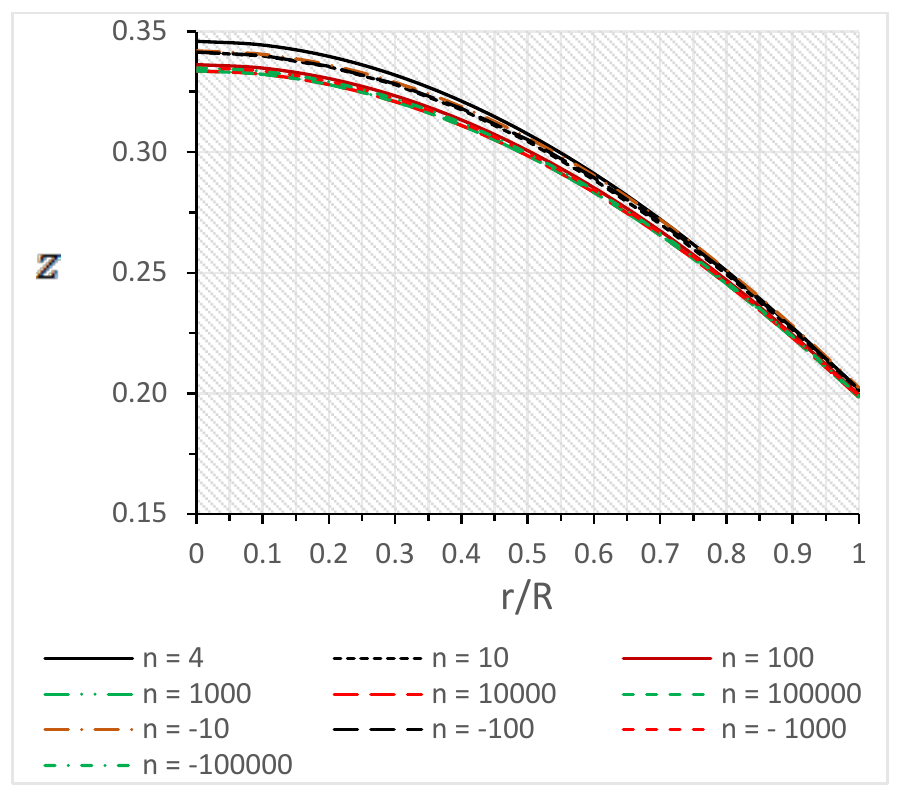}
\caption{Behavior of redshift ($z$) verses fractional radius $r/R$ for Her X-1.The numerical values of the constants are given in tables 1 $\&$ 2.}
\label{energy1}
\end{figure}

\begin{table}
\centering \caption{Numerical values of physical parameters $AR^2$, $A$, $C$, $B$, $E_0$ for the compact star Her X-1 of mass $M=0.85\left(M_\odot\right)$, radius $R=8.10~(km)$ ~\cite{Gangopadhyay} for the negative values of $n$} \label{Table1}

{\begin{tabular}{@{}c|ccccccc@{}}

\hline

$n$ & $AR^2$ & $A$ & $C$ & $B$& $K$ & $E_0$  \\ \hline

$-10$ & -0.021727 & -0.3311$\times10^{-3}$ & -15.7670 & 0.6676 & 7.1328$\times10^{2}$ & 0.0200 $\times10^{2}$ \\

$-100$ & -0.002200 & -0.3350$\times10^{-4}$ & -1.6121$\times10^{2}$ & 0.66768 & 7.1328$\times10^{2}$& 0.0347$\times10^{4}$ \\

$-1000$ & -0.000215 & -0.3275$\times10^{-5}$ & -1.6407$\times10^{3}$ & 0.67267 &7.2074$\times10^{2}$ & 0.0906$\times10^{6}$ \\

$-10000$ & -0.0000215 & -0.3275$\times10^{-6}$ & -1.6411$\times10^{4}$ & 0.67272 & 7.4475$\times10^{2}$ & 0.0906 $\times10^{8}$ \\

$-100000$ & -0.00000215 & -0.3275$\times10^{-7}$ & -1.6411$\times10^{5}$ & 0.67271 &7.4489$\times10^{2}$ & 0.0906 $\times10^{10}$ \\ \hline
\end{tabular}}
\end{table}
\begin{table}
\centering \caption{Numerical values of physical parameters $AR^2$, $A$, $C$, $B$, $E_0$ for the compact star Her X-1 of mass $M=0.85\left(M_\odot\right)$, radius $R=8.10~(km)$ ~\cite{Gangopadhyay} for the positive values of $n$} \label{Table2}

{\begin{tabular}{@{}c|ccccccc@{}}

\hline

$n$ & $AR^2$ & $A$ & $C$ & $B$ & $K$ &  $E_0$  \\ \hline

$4$ & 0.05836 & 0.889$\times10^{-3}$ & 6.79354 & 0.6632 & 7.2012$\times10^{2}$  & 0.48 \\

$10$ & 0.022375 & 0.341$\times10^{-3}$ & 16.5571 & 0.6674 & 7.2752$\times10^{2}$  & 0.0500$\times10^{2}$ \\

$100$ & 0.002164 & 0.3298$\times10^{-4}$ & 1.6399$\times10^{2}$ & 0.6717 &7.4030$\times10^{2}$ & 0.0805$\times10^{4}$ \\

$1000$ & 0.000214 & 0.3261$\times10^{-5}$ & 1.6466$\times10^{3}$ & 0.67368 &7.4953$\times10^{2}$  & 0.1026$\times10^{6}$ \\

$10000$ & 0.000021395 & 0.3260$\times10^{-6}$ & 1.6464$\times10^{4}$ & 0.6737 & 7.4962$\times10^{2}$  & 0.1027 $\times10^{8}$ \\

$100000$ & 0.0000021394 & 0.3260$\times10^{-7}$ & 1.6463$\times10^{5}$ & 0.6737 &7.4960$\times10^{2}$  & 0.1027$\times10^{10}$ \\  \hline
\end{tabular}}
\end{table}

\begin{table}
\centering \caption{Physical Parameters for Her X-1 for negative values of $n$}\label{Table3}

{\begin{tabular}{@{}c|ccccccc@{}}

\hline

value & Central Density & Surface Density & Central Pressure & Surface  \\
of $n$ & $gm/cm^{3} $ & $gm/cm^{3}$ & $dyne/cm^{2}$ & Redshift \\ \hline

$-10$ &8.41017 $\times10^{14}$ & 6.93513 $\times10^{14}$ & 6.77501 $\times10^{34}$ & 0.202452 \\

$-100$ &8.70023 $\times10^{14}$ & 6.76264$\times10^{14}$ & 6.28172 $\times10^{34}$ & 0.201663  \\

$-1000$ & 8.65631$\times10^{14}$ & 6.65753$\times10^{14}$ & 5.68838 $\times10^{34}$ & 0.198982\\

$-10000$ &8.65819 $\times10^{14}$ & 6.65540 $\times10^{14}$ & 5.68276$\times10^{34}$ & 0.198964  \\

$-100000$ & 8.65837$\times10^{14}$ & 6.65519 $\times10^{14}$ & 5.68220 $\times10^{34}$ & 0.198943 \\ \hline
\end{tabular}}
\end{table}

\begin{table}
\centering \caption{Physical Parameters for Her X-1 for positive values of $n$} \label{Table3}

{\begin{tabular}{@{}c|ccccccc@{}}

\hline

value & Central Density & Surface Density & Central Pressure & Surface  \\
of $n$ & $gm/cm^{3} $ & $gm/cm^{3}$ & $dyne/cm^{2}$ & Redshift \\ \hline

$4$ & 9.72956 $\times10^{14}$ & 6.36703$\times10^{14}$ & 5.18467$\times10^{34}$ & 0.201706  \\

$10$ & 9.09566 $\times10^{14}$ & 6.57299 $\times10^{14}$ & 5.67525 $\times10^{34}$ & 0.200921  \\

$100$ & 8.71270$\times10^{14}$ &6.65970$\times10^{14}$ & 5.74156$\times10^{34}$ & 0.199415  \\

$1000$ &8.65050$\times10^{14}$ & 6.63819$\times10^{14}$ & 5.57044$\times10^{34}$ & 0.198444 \\

$10000$ & 8.64653$\times10^{14}$ & 6.63888$\times10^{14}$ & 5.57271$\times10^{34}$ & 0.198438 \\

$100000$ & 8.64642$\times10^{14}$ & 6.63921 $\times10^{14}$ & 5.57305 $\times10^{34}$ & 0.198433 \\ \hline
\end{tabular}}
\end{table}

\begin{table}[h]
\centering
\caption{List of embedding class one solutions with well behaved nature of $dp_i/d\rho$ for the {\it ansatz} $e^{\nu(r)} = B(1+A\,r^2)^{n}$}\label{tbl-4}
\begin{tabular}{@{}lrrrrrrrrr@{}}
\hline
 $n$ \& $A$                   & Electric charge           & Pressure                     & Well behaved                      & Reference\\
                              & function ($E$)            &  anisotropy ($\Delta $)      & nature of $dp_i/d\rho$                    &\\  \hline

$n=2, A \ge0 $                & $E\ne0$                  & $\Delta \ne 0$ (EOS)          &    No                             & ~\cite{Piyali1}\\
$n=4, A \ge0 $                & $E=0$                    & $\Delta \ne 0$                &    yes                             & ~\cite{Piyali2}\\ \hline

$n, A \in \Re^{+} \cup 0 $    & $E=0$                      & $\Delta \ne 0$              &    Yes~($n \geq 3$)               & ~\cite{MGSD1}\\

$n,A \in \Re^{+}\cup 0  $    & $E \ne 0$                   &  $\Delta =0 $              &    Yes~($n \geq 3.3$)             & ~\cite{MGSD2}\\ \hline

$n, A \in \Re^{-}\cup 0 $    & $E=0$                       & $\Delta \ne 0$              &    Yes~($n \leq -3$)               & ~\cite{MDSP}\\

$n, A \in \Re^{-}\cup 0  $   & $E \ne 0$                   &  $\Delta =0 $            &    Yes~($n \leq -2.7$)             & ~\cite{sungov}\\ \hline

$n, A \in \Re $          & $E=E_0A^2r^4\,(1+Ar^2)^n$      &  $\Delta \ne 0 $      &    Yes                                        & ~Present case\\
                              &                             &                      &  $n \in (-\infty, -3] \cup [2.7, \infty) $     & \\  \hline

\end{tabular}
\end{table}

\section{Discussion of results}

We have presented an exact static model of the Einstein-Maxwell equations which describes a spherically symmetric charged body arising from the requirement that the internal geometry is of embedding class I. The energy momentum tensor describes an anisotropic fluid with an electromagnetic field. Fig. 1 displays the trend in the gravitational potentials as a function of the dimensionless ratio, $r/R$. The gravitational potentials are continuous and increase smoothly from the center of the star towards the surface. An increase/decrease in $n$ has no appreciable effect in the magnitude or nature of the gravitational potentials. In Fig. 2 we present the trend in the charge (left panel) and the mass (right panel). Both the charge and the mass vanish at the centre of the configuration and increases monotonically towards the surface of the star. An increase in $|n|$ is accompanied by an increase in both the charge and the mass. The divergence is greater towards the surface layers of the star. Fig. 3 illustrates the behaviour of the radial pressure (left panel) and the tangential pressure (right panel). It is clear that the radial and tangential pressures are monotonically decreasing functions of the radial coordinate. The radial pressure vanishes at some finite radius which defines the boundary of the star. We note that the tangential pressure is nonvanishing at the stellar surface. We also note that the radial and tangential pressures increase with an increase in $|n|$. This increase is noticeable closer to the inner layers of the star and is indistinguishable as the surface layers are approached. A very large increase in $|n|$ of the order of $10^3$ has very little effect on the relative magnitudes of both the radial and transverse stresses throughout the interior of the star. The trend of the density is profiled in fig. 4. We observe that the density is a monotonically decreasing function attaining a maximum value at the centre of the star. We observe an interesting trend in the density as $|n|$ increases. An increase in $|n|$ is accompanied by a decrease in the density at each interior point of the gravitating body. For very large values of $|n|$ the density profile is approximately the same for each interior point.  The decrease in the density is most noticeable closer to the center of the star. The anisotropy parameter is displayed in Fig. 5. As pointed earlier, the anisotropy parameter vanishes at the centre of the star and increases monotonically outwards towards the surface. An increase $|n|$ is accompanied by an increase in $\Delta$ with the relative differences being more marked towards the surface layers of the star. All the energy conditions are satisfied at each interior point of the configuration as displayed in fig. 6. The various forces operating within the stellar interior are plotted in fig. 7. It has been pointed out that a change in $|n|$ effects changes in the force due to anisotropy and the electromagnetic force. For small values of $|n|$ we pointed out that the anisotropic force dominates the electromagnetic force. This trend switches over for large $|n|$. Our model obeys the causality condition throughout the stellar interior (Fig.8).The stability of our model was studied by looking at the relative sound speeds squared (Fig. 9). The Abreu et al. stability analysis shows that there are no unstable regions within the stellar core indicating that the likelihood of cracking occurring within our model is remote (Fig. 10). Fig. 7 reveals a new phenomenon associated with these models. We note that for the first time that the relative difference between the electromagnetic force and the force due to anisotropy can change sign and this is directly related to an increase in $|n|$. Figure 11 shows the trend of the redshift inside the star. The details of the embedding class one solutions with well-behaved nature of $d pi /d\rho$ for the ansatz (27) $\nu(r ) = B\,(1 + A\,r^2)^n $ is given by Table 5.\\

{\bf Acknowledgements} The author S. K. Maurya acknowledges authority
of University of Nizwa for continuous support and encouragement to
carry out this research work.


\end{document}